\documentclass[11pt,reqno]{amsart}  

\usepackage{amscd} 
\usepackage{mathrsfs}
\usepackage{amsmath,amsbsy,amsthm,amsfonts,amssymb,esint}
\usepackage[unicode,psdextra]{hyperref}
\usepackage{graphicx}
\usepackage{tikz}
\usepackage{bm}
\usepackage{soul}
\usepackage{color}

\usepackage[sort&compress,capitalise,nameinlink]{cleveref}
\crefname{section}{\textsection}{\textsection}
\crefname{subsection}{\textsection}{\textsection}
\usepackage{dsfont}
\usepackage{enumerate}

\DeclareMathAlphabet\mathbfcal{OMS}{cmsy}{b}{n}

\newtheorem{theorem}{Theorem}[section]

\newtheorem{proposition}[theorem]{Proposition}
 
\theoremstyle{definition}

\newtheorem{remark}[theorem]{Remark} 
\numberwithin{equation}{section}

\def\A{\mathbf{A}}
\def\B{\mathbf{B}}
\def\R{\mathbb{R}}
\def\Z{\mathbb{Z}}
\def\C{\mathbb{C}}
\def\x{\bm{x}}
\def\0{\bm{0}}
\def\fb{\bm{f}}
\def\Zpt{\mathcal{Z}_{\Pi,\Theta}}
\def\Ha{H_{\alpha}}
\def\Hat{\mathring{H}_{\alpha}}
\def\dHa{\bar{H}_{\alpha}}
\def\hal{h_{\alpha,\ell}}
\def\dhal{\bar{h}_{\alpha,\ell}}
\def\dhals{\bar{h}_{\alpha,\ell}^{*}}
\def\Aa{\A_{\alpha}}
\def\dom{\mathcal{D}}
\def\ran{\mbox{ran}}
\def\KK{\mathcal{K}}
\def\GG{G}
\def\Gl{G}
\def\UU{\hat{U}}
\def\cIl{c_{I,\ell}}
\def\cKl{c_{K,\ell}}
\def\vl{v_{\ell}}
\def\wl{w_{\ell}}
\def\Qa{Q_{\alpha}}
\def\QaF{Q_{\alpha}^{(\mathrm{F})}}
\def\HaF{H_{\alpha}^{(\mathrm{F})}}
\def\HaK{H_{\alpha}^{(\mathrm{K})}}
\def\RaF{R_{\alpha}^{(\mathrm{F})}}
\def\GG{\mathcal{G}}
\def\GGc{\breve{\GG}}
\def\tt{\bm{\tau}}
\def\bb{\bm{\beta}}
\def\H0{H_0}

\def\k{\bm{k}}
\def\Fou{\mathfrak{F}}

\def\Su{\mathbb{S}^1}

\DeclareMathOperator*{\slim}{s\,--\,lim}

\NewCommandCopy{\mlmhbar}{\hbar}
\NewCommandCopy{\amshbar}{\hbar}
\RenewCommandCopy{\hbar}{\mlmhbar}
\def\hb{\amshbar}

\title[The Aharonov-Bohm Hamiltonian: self-adjointness, spectral and scattering properties]{The Aharonov-Bohm Hamiltonian: self-adjointness, spectral and scattering properties}

\author{Davide Fermi}
\address{Dipartimento di Matematica, Politecnico di Milano, P.zza Leonardo da Vinci, 32, 20133, Milano, Italy\\
and Istituto Nazionale di Fisica Nucleare, Sezione di Milano, Italy}
\email{davide.fermi@polimi.it}
\urladdr{https://fermidavide.com}

\begin{document}

\begin{abstract}
This work provides an introduction and overview on some basic mathematical aspects of the single-flux Aharonov-Bohm Schr\"odinger operator. The whole family of admissible self-adjoint realizations is characterized by means of four different methods: von Neumann theory, boundary triplets, quadratic forms and Kre{\u\i}n's resolvent formalism. 
The relation between the different parametrizations thus obtained is explored, comparing the asymptotic behavior of functions in the corresponding operator domains close to the flux singularity.
Special attention is devoted to those self-adjoint realizations which preserve the same rotational symmetry and homogeneity under dilations of the basic differential operator. The spectral and scattering properties of all the Hamiltonian operators are finally described.
\end{abstract}


\maketitle

\section{Introduction}
It is nowadays a well-established fact that electromagnetic fields affect the dynamics of quantum particles, even when the latters are confined to regions where the fields are absent. In this context, even though there is no force acting on the particles in the classical sense, the wave-function experiences an observable phase shift, which can be expressed in terms of the corresponding electromagnetic potentials. The theoretical prediction of this phenomenon was first outlined in \cite{ES49}, but its fundamental significance in Quantum Mechanics was brought to the fore a decade later by the independent derivation presented in \cite{AB59}, after which it is named.

An early claim of experimental observation of the Aharonov-Bohm effect (AB effect) was first reported in \cite{Ch60}, while undisputed evidence of the associated interference patterns was later provided in \cite{TO86}; see also \cite{PT89} and \cite{BT09}. On the other hand, the AB effect has been for years at the center of an interpretation debate, questioning the locality principle and the physical reality of electromagnetic potentials \cite{AB61,V12,K15}. Recent investigations indicate that a sensible explanation in terms of local fields interactions can actually be attained via a full-fledged Quantum Electrodynamics description \cite{MV20,K22}.
It is also worth mentioning that AB-like phenomena play a major role in different scenarios of condensed matter physics, including: {\it anyonic excitations} of 2D electron gases in the Fractional Quantum Hall Effect \cite{LM77,Wi82,BK20,NL20}; magnetic vortexes appearing in planar samples of {\it type-II superconductors} \cite{Ab57}; the localization of non-interacting particles in {\it AB cages} \cite{BPC21,MVV22}.

From the mathematical perspective, even the simplest AB model displays a rich structure and a number of interesting features. The archetypal configuration examined in \cite{AB59} considers a scalar particle in presence of a straight ideal solenoid. This is basically described by a magnetic Schr\"odinger operator comprising a vector potential with a strong local singularity and a long-range tail, as well. The rigorous analysis of the self-adjointness, spectral and scattering properties of this operator was first addressed in \cite{Ru83} and further extended in \cite{AT98,DS98,RY02,Ya03,Ya06,PR11,CO18,Y21}. The presence of additional regular magnetic perturbations, on top of the AB singularity, was considered in \cite{ESV02,CF21,F24}. Schr\"odinger operators with more than one AB flux were studied in \cite{St89,St91,KW94,DFO95,DFO97,Na00,IT01,MOR04,Ta07,Ta08,FG08,BMO10,AT11,AT14,CF24a,CF24b}. The approximation of AB Schr\"odinger Hamiltonians by means of regular electrostatic and magnetic potentials was examined in \cite{MVG95,Ta01,OP08}. Classical results on Hardy-type inequalities and dispersive estimates for the single-flux Schr\"odinger operator can be found in \cite{LW99,GK14,BDELL20}. The spectral properties of Hamiltonians with an AB singularity in compact domains were studied, \emph{e.g.}, in \cite{AFNN17,AN18,FNOS23}.
Pauli and Dirac Hamiltonians describing the motion of a spin $1/2$ particle in presence of singular magnetic potentials of AB-type were analyzed in \cite{EV02,Ta03,GS04,Pe05,Pe06,KA14,BCF}.

In the present work we discuss the fundamental mathematical aspects of the original single-flux AB Schr\"odinger operator. We begin by describing the self-adjoint realizations of this operator, employing four distinct methodologies and exploring their interconnections. Specifically, we characterize the complete family of self-adjoint realizations using von Neumann theory, boundary triplets, quadratic forms, and Kre{\u\i}n's resolvent methods. Next, we outline the primary spectral and scattering properties of the corresponding Hamiltonian operators.

\section{The Aharonov-Bohm model}

The prototypical AB setting considers a charged quantum particle moving outside of a straight cylindrical solenoid. The attention is restricted to the low energy regime where the de Broglie wavelength of the particle is much larger than the solenoid diameter and, at the same time, much smaller than the longitudinal length of the solenoid itself. To leading order, this model is described in terms of an ideal solenoid of zero diameter and infinite length, matching the singular magnetic field
\begin{equation}\label{eq:BAB3d}
	\B(\x) = \Phi\,\delta_{(x,y) = (0,0)}\, \bm{e}_z\,.
\end{equation}
Here, $\x = (x,y,z) \in \R^3$ is a system of Cartesian coordinates, $\Phi \in \R$ is the total magnetic flux, $\delta_{(x,y) = (0,0)}$ is the 2D Dirac delta distribution concentrated on the $z$-axis and $\bm{e}_z$ is the unit vector parallel to the same axis. 

An admissible vector potential associated to the magnetic field \eqref{eq:BAB3d} can be firstly inferred by a heuristic calculation, based on Helmholtz's decomposition theorem \cite[\S 3.9]{AWH13}:
\begin{align*}
	& \A(\x) = \frac{1}{4\pi} \int_{\R^3} d\x'\; \frac{\nabla'\! \wedge \B(\x')}{|\x - \x'|}
	\,=\, -\, \frac{1}{4\pi} \int_{\R^3} d\x'\; \nabla' \!\left(\tfrac{1}{|\x - \x'|}\right) \wedge \B(\x') \\
	& = -\, \frac{\Phi}{4\pi} \int_{\R} dz'\; \frac{(x,y,z - z') \wedge \bm{e}_z }{[x^2+y^2 + (z-z')^2]^{3/2}} 
	\,=\, \frac{\Phi}{2\pi}\left(-\,\frac{y}{x^2+y^2}\,,\, \frac{x}{x^2+y^2}\,,\,0\right).
\end{align*}
We take the latter expression as the very definition of $\A$. Then, working backward, it can be rigorously proved that, in the sense of Schwartz's distributions,
\begin{equation*}
	\nabla \wedge \A = \B\,.
\end{equation*}
Indeed, for any smooth and compactly supported vector field $\fb \in C^{\infty}_c(\R^3;\R^3)$, integrating by parts in the cylindrical region outside the $z$-axis, we get
\begin{align*}
  & \langle \nabla \wedge \A, \fb \rangle 
    \,=\, \langle \A, \nabla \wedge \fb \rangle 
    \,=\, \frac{\Phi}{2\pi} \lim_{\varepsilon \to 0^+} \int_{\{x^2+y^2 \,>\, \varepsilon^2\}} d\x\;\frac{(-y,x,0)}{x^2+y^2} \cdot (\nabla \wedge \fb) \\
  & =\, \frac{\Phi}{2\pi} \lim_{\varepsilon \to 0^+} \left[
  		- \varepsilon \int_{\{x^2+y^2 \,=\, \varepsilon^2\}} d\sigma\;\frac{(-y,x,0) \cdot [(x,y,0) \wedge \fb]}{(x^2+y^2)^{3/2}} \right. 
  		\\
  		& \hspace{3cm} \left. + \int_{\{x^2+y^2 \,>\, \varepsilon^2\}} d\x\;\fb \cdot \nabla \wedge \left(-\,\frac{y}{x^2+y^2}\,,\, \frac{x}{x^2+y^2}\,,\,0\right)
  		\right] \\
  & =\, \frac{\Phi}{2\pi} \lim_{\varepsilon \to 0^+} \int_{\{x^2+y^2 \,=\, \varepsilon^2\}} d\sigma\;\bm{e}_z \cdot \fb
  \;=\; \langle \Phi\,\delta_{(x,y) = (0,0)}\, \bm{e}_z\,,\,\fb \rangle\,.
\end{align*}
A similar computation shows that $\A$ fulfills the Coulomb gauge, again in the sense of distributions, \emph{i.e.},
\begin{equation}
	\nabla \cdot \A = 0\,.
\end{equation}

On account of the above premises, the one-particle quantum dynamics for the AB configuration is described by means of the three-dimensional Schr\"odinger operator
\begin{equation*}
	H_{\mathrm{3D}} = \frac{1}{2m}\, \big(-i \hb \nabla - q \,\A\big)^2\,,
\end{equation*}
where $\hb$ is the reduced Plank constant, $m$ is the mass of the particle, $q$ its electric charge and $\A$ is the vector potential introduced above.

\subsection{The reference Schr\"odinger operator}
The previously described setting is manifestly invariant under translations along the $z$-axis, see \eqref{eq:BAB3d}. Based on this, in the sequel we focus our attention on the two-dimensional problem obtained by factoring out the axial direction and {\it effectively suppressing the $z$-coordinate}. After transition to natural units of measure via the rescaling $\x \mapsto \frac{\hb}{\sqrt{2m}}\,\x$, we shall henceforth exclusively refer to the two-dimensional Schr\"odinger operator
\begin{equation}\label{eq:Hadef}
	\Ha \doteq (-i \nabla + \Aa)^2\,,
\end{equation}
depending on the flux parameter $\alpha = - \frac{q \Phi}{2\pi \hb}$ through the reduced AB potential
\begin{equation}\label{eq:Aadef}
	\Aa(\x) \doteq \alpha\,\frac{\x^{\perp}}{|\x|^2}  \qquad (\x \in \R^2)\,.
\end{equation}
Let us stress that here and in the remainder of this work, by a slight abuse of notation, we refer to two-dimensional Cartesian coordinates $\x = (x,y) \in \R^2$, setting $\nabla = (\partial_x,\partial_y)$ and $\x^{\perp} = (-y,x)$ for brevity.

Given the singular behavior of $\Aa$ at $\x = \0$, we initially regard $\Ha$ as a densely defined operator in $L^2(\R^2)$ with domain given by the set of smooth functions with compact support away from the flux position, namely,
\begin{equation}\label{eq:domHa}
	\dom(\Ha) \doteq C^{\infty}_c(\R^2 \setminus \{\0\})\,.
\end{equation}
With this understanding, $\Ha$ is certainly symmetric, whence closable. Yet, it is not closed, let alone self-adjoint. The construction of self-adjoint extensions of $H_{\alpha}$ will be addressed in the next section.

Before proceeding, let us point out that it entails no loss of generality to assume
\begin{equation}\label{eq:al01}
	\alpha \in (0,1)\,.
\end{equation}
To justify this claim, consider the maps $(U_k \psi)(\x) =  e^{i k \arg \x} \, \psi(\x)$ ($k \in \Z$), where $\arg \x = \arctan(y/x)$ is any determination of the argument (\emph{viz.}, angle) of $\x \in \R^2 \setminus \{\0\}$. It is easy to see that such $U_k$ are well-defined and identify unitary operators in $L^2(\R^2)$. Noting that $\nabla \arg \x = \frac{\x^{\perp}}{|\x|^2}$, a direct computation yields $U_k \,\Ha\, U_{k}^{-1} = H_{\alpha - k}$, thus proving that $\Ha$ and $H_{\alpha - k}$ are unitarily equivalent for any $k \in \Z$. 

The range of the flux parameter could be further restricted exploiting the conjugation symmetry (which actually coincides with the reflection $z \mapsto -z$ in the three-dimensional formulation). As a matter of fact, setting $(C \psi)(x) = \psi^{*}(x)$, one can readily check that $C\, H_{\alpha}\, C^{-1} = H_{-\alpha}$. So, combining $C$ with one of the previous maps $U_k$, it is always possible to fix $\alpha \in (0,1/2]$ via a anti-unitary transformation. Throughout this work we shall leave aside this option and stick to the weaker requirement \eqref{eq:al01}, thus avoiding anti-linearity issues.

\begin{remark}[False impressions]
The operators $U_k$ mentioned below Eq. \eqref{eq:al01} do not correspond to any meaningful gauge transformation, since their action actually modifies the flux parameter and, thus, the magnetic field. 
An analogous misconception regards the possibility to simply ``gauge away'' the Aharanov-Bohm flux, using the map $(U_\alpha \psi)(\x) = e^{i \alpha \arg \x} \, \psi(\x)$. As a matter of fact, a naive algebraic calculation yields $U_\alpha \,\Ha\, U_{\alpha}^{-1} = -\Delta$, where  $-\Delta$ is the Laplace differential operator. However, it should be noticed that $U_\alpha \psi$ is not single-valued, so $U_{\alpha}$ does not define any proper operator in $L^2(\R^2)$. A precise formulation would require treating $U_\alpha \psi$ as a function on a Riemann surface and, accordingly, identifying $U_\alpha \,\Ha\, U_{\alpha}^{-1}$ with the free Laplace operator acting thereon.
This approach is indeed tenable (see, \emph{e.g.}, \cite{St89,St91,Na00,GS04}), but we shall not discuss it any further in this chapter for the sake of a clearer presentation.
\end{remark}

\begin{remark}[Invariance under rotations and homogeneity under dilations]\label{rem:homdil}
Expressions \eqref{eq:Hadef} and \eqref{eq:Aadef} make evident that the differential operator $\Ha$ is invariant under rotations and homogeneous of degree $-2$ under dilations. To be more precise, let us introduce the unitary groups $\{T_{R}\}_{R \,\in\, \mathrm{SO}(2)}$ ($\mathrm{SO}(2)$ is the special orthogonal group in dimension $2$) and $\{D_{\gamma}\}_{\gamma \,\in\, \R}$ acting in $L^2(\R^2)$, respectively, by
\begin{equation}\label{eq:dilat}
	(T_{R} \psi)(\x) = \psi(R \x)\,, \qquad 
	(D_{\gamma} \psi)(\x) = e^{\gamma}\;\psi(e^{\gamma}\,\x)\,.
\end{equation}
By direct inspection, we readily obtain
\begin{equation*}
	T_{R} \,\Ha\, T_{R}^{-1} = \Ha\,, \qquad
	D_{\gamma} \,\Ha\, D_{\gamma}^{-1} = e^{-2\gamma}\, \Ha\,,
\end{equation*}
alongside with the trivial identities $T_{R}^{-1} \dom(\Ha) = \dom(\Ha)$ and $D_{\gamma}^{-1} \dom(\Ha) = \dom(\Ha)$, see \eqref{eq:domHa}.
We shall show in the sequel that, apart from notable exceptions, these invariance and homogeneity properties are generically violated by the self-adjoint extensions of $\Ha$, due to domain-related issues.
Incidentally, we also mention that the parity operator $(P \psi)(x,y) = \psi(-x,y)$ is such that $P \Ha P^{-1} = H_{-\alpha}$.
\end{remark}

\section{Self-adjoint realizations}
We formerly introduced the fundamental Schr\"odinger operator $\Ha$, regarding it as a densely defined symmetric operator, see \eqref{eq:Hadef} and \eqref{eq:domHa}. In this section we classify all the admissible self-adjoint extensions of $\Ha$ in $L^2(\R^2)$, thereby identifying different Hamiltonian operators for the AB model. We describe four different approaches based on von Neumann theory, boundary triplets, quadratic forms and Kre{\u\i}n's theory, respectively. The connection between the different parametrizations of the self-adjoint realizations thus obtained will be discussed along the way.

\subsection{The von Neumann construction}
The most direct way to obtain all the self-adjoint extensions of $\Ha$ is perhaps through standard von Neumann theory, computing exactly the related deficiency subspaces. A fully explicit analysis can be indeed conducted by leveraging on the rotational symmetry of the model, which allows for a partial-wave decomposition. Hereafter, we outline this approach, essentially retracing the original arguments presented in \cite{AT98}.

As a starting point, we refer to the closure of $\Ha$, which we denote by $\dHa$. Its adjoint $\dHa^{*}$ is defined on the domain
\begin{equation}\label{eq:domHas}
	\dom\big(\dHa^{*}\big) = \big\{\psi \in L^2(\R^2)\,\big|\, \psi \in H^2_{\mathrm{loc}}(\R^2 \setminus \{\0\})\,,\; \Ha \psi \in L^2(\R^2)\big\}\,,
\end{equation}
where, $H^2_{\mathrm{loc}}(\R^2 \setminus \{\0\})$ is the local Sobolev space of order two.
Keeping this in mind, we now inspect the deficiency equation
\begin{equation*}
	\dHa^{*}\, G_{\pm} = \pm i\,G_{\pm}\,.
\end{equation*}

To this purpose, we pass to a set of polar coordinates $(r,\theta) \in \R_{+}\! \times \Su$, with radius $r = |\x|$ and angle $\theta = \arg \x$ belonging to the circle $\Su = \R / 2\pi\,\Z \simeq [0,2\pi)$. We consider the corresponding decomposition
\begin{equation*}
	L^2(\R^2) = L^2(\R_{+},\,r\,dr) \otimes L^2(\Su,d\theta)\,,
\end{equation*}
and further notice that, as a differential operator, $\Ha$ can be expressed as
\begin{equation*}
	\Ha = - \,\tfrac{1}{r}\,\partial_r (r\,\partial_r \;\cdot\,) + \tfrac{1}{r^2}\, (-i \partial_\theta + \alpha)^2\,.
\end{equation*}
Introducing the unitary map
\begin{equation*}
	V : L^2(\R_{+},\,r\,dr) \to L^2(\R_{+},dr)\,, \qquad (V f)(r) = \sqrt{r}\,f(r)\,,
\end{equation*}
and exploiting the completeness of the angular harmonics $\left\{\tfrac{e^{i \ell \theta}}{\sqrt{2\pi}}\right\}_{\ell \in \Z}$ in $L^2(\Su,d\theta)$, we derive the following representations:
\begin{gather*}
	L^2(\R^2) = \bigoplus_{\ell \in \Z}\; V^{-1} L^2(\R_{+},dr) \,\otimes\, \mathrm{span}\left( \tfrac{e^{i \ell \theta}}{\sqrt{2\pi}} \right) \,; \\
	\dHa = \bigoplus_{\ell \in \Z}\; V^{-1}\, \dhal\, V \,\otimes\, \bm{1}\,.
\end{gather*}
Here, for $\ell \in \Z$ and $\alpha \in (0,1)$, see \eqref{eq:al01}, we have indicated with $\dhal$ the closure of the radial differential operator
\begin{equation}\label{eq:hal}
	\hal \doteq -\,\frac{d^2}{dr^2} + \frac{(\ell+\alpha)^2 - 1/4}{r^2}\,,
\end{equation}
initially defined on $C^{\infty}_c(\R_{+})$. Explicit characterizations of the domains $\dom\big( \dhal \big)$ can be found in \cite{AT98}. It is noteworthy that $\dhal$ commutes with the complex conjugation, so it certainly admits self-adjoint extensions, see \cite[Vol. II, Thm. X.3]{RS}. 
To determine such extensions, we consider the adjoint operator $\dhals$, with domain
\begin{equation}\label{eq:domhals}
	\dom\big(\dhals\big) = \big\{\xi \in L^2(\R_{+},\,dr)\;\big|\; \xi \in H^2_{\mathrm{loc}}(\R_{+})\,,\; \hal \xi \in L^2(\R_{+},\,dr)\big\}\,,
\end{equation}
and proceed to examine the associated radial deficiency equation
\begin{equation}\label{eq:defradeq}
	\dhals\, \xi^{(\ell)}_{\pm} = \pm i\,\xi^{(\ell)}_{\pm}\,.
\end{equation}
The only admissible solutions of the above equation are of the form
\begin{equation*}
	\xi^{(\ell)}_{\pm}(r) = \cIl\, \sqrt{r}\; I_{|\ell + \alpha|}\big(e^{\mp i \frac{\pi}{4}} r\big) + \cKl\, \sqrt{r}\;K_{|\ell + \alpha|}\big(e^{\mp i \frac{\pi}{4}} r\big)\,,
\end{equation*}
where $\cIl,\cKl \in \C$ are suitable constants and $I_{\nu},K_{\nu}$ are the standard modified Bessel functions of the second kind, see \cite[\S 10.25]{NIST}. For $\alpha \in (0,1)$, considering the regularity features of these functions and their asymptotic behavior for $r \to +\infty$ and $r \to 0^+$, see \cite[\S 10.30]{NIST}, it appears that the basic condition $\xi^{(\ell)}_{\pm} \in L^2(\R_{+},dr)$ in \eqref{eq:domhals} can be realized if and only if $\cIl = 0$ for all $\ell \in \Z$ and $\cKl = 0$ for all $\ell \in \Z \setminus \{0,-1\}$.
It follows from here that the deficiency indexes of $\dhal$ are as follows: $(0,0)$ (\emph{i.e.}, $\dhal$ is essentially self-adjoint) for all $\ell \in \Z \setminus \{0,-1\}$; $(1,1)$ for $\ell \in \{0,-1\}$.
For a finer analysis of the self-adjointness and spectral properties of the radial operator \eqref{eq:hal}, we refer to \cite{BDG11,DR17,DG21,DFNR20,DF23}.

Building on the above arguments, it can be inferred that $\dHa$ has deficiency indexes $(2,2)$, with deficiency subspaces $\KK_{\pm}$ spanned by
\begin{equation}\label{eq:chiell}
	G^{(\ell)}_{\pm}(r,\theta) \doteq e^{\mp i \frac{\pi}{4}\, |\ell+\alpha|}\;K_{|\ell+\alpha|}\big(e^{\mp i \frac{\pi}{4}}\, r\big)\, \tfrac{e^{i \ell \theta}}{\sqrt{2\pi}}
	\qquad \big(\ell \in  \{0,-1\}\big)\,.
\end{equation}
Adopting a slightly improper terminology, we will refer to $G^{(0)}_{\pm}$ and $G^{(-1)}_{\pm}$ as {\it s-wave} and {\it p-wave defect functions}, respectively.
The specific choice of the overall phases in \eqref{eq:chiell}, seemingly arbitrary at this stage, will be justified {\it a posteriori} by the simpler parametrization of some notable self-adjoint extensions of $\dHa$ (see next Remarks \ref{rem:vNhom}, \ref{rem:asyF} and \ref{rem:asyQuad}). One could even introduce a normalization constant in the above expression for $G^{(\ell)}_{\pm}$, yet this is immaterial for our purposes.

For convenience of presentation, let us now introduce the maps
\begin{equation}\label{eq:defGpm}
	\GG_{\pm} : \C^2 \to L^2(\R^2)\,, \qquad \GG_{\pm}\, \bm{c} \doteq \mbox{$\sum_{\ell\,\in\,\{0,-1\}}$}\,c_{\ell}\;G^{(\ell)}_{\pm}\,,
\end{equation}
and parametrize the deficiency subspaces as
\begin{equation*}
	\KK_{\pm} \doteq \ker\big(\dHa^{*}\mp i\big)
	= \ran\, \GG_{\pm} = \big\{\GG_{\pm} \bm{c} \in L^2(\R^2)\;\big|\;\bm{c} = (c_{0},c_{-1}) \!\in \C^2\big\}\,.
\end{equation*}
It goes without saying that $\GG_{\pm} : \C^2 \to \KK_{\pm}$ are continuous bijections. Moreover, any unitary operator $\UU : \KK_{+} \to \KK_{-}$ is uniquely represented by a unitary matrix $U \in \mathrm{U}(2)$ ($\mathrm{U}(2)$ is the unitary group of degree $2$), such that
\begin{equation}\label{eq:UGGU}
	\UU \,\GG_{+} = \GG_{-}\, U \,.
\end{equation}

Summing up, a direct application of von Neumann theory \cite[Vol. II, Thm. X.2]{RS} ultimately implies the following \cite{AT98}.

	\begin{theorem}[Self-adjoint extensions - von Neumann theory]\label{thm:vNext}
		Let $\alpha \in (0,1)$. Then, the self-adjoint extensions on $L^2(\R^2)$ of the closed symmetric operator $\dHa$ are in one-to-one  correspondence with the unitary matrices $U \in \mathrm{U}(2)$. Moreover, for any $U\in \mathrm{U}(2)$, the domain and action of the self-adjoint extension $\Ha^{(U)}$ are, respectively,
		\begin{gather}
		\dom\big(\Ha^{(U)}\big) = \big\{ \psi = \varphi + \GG_{+} \bm{c} + \GG_{-} U \bm{c} \,\in L^2(\R^2) \;\big|\; \varphi \in \dom(\dHa)\,,\; \bm{c}\!\in \C^2 \big\}\,, \label{eq:vNdom} \\
		\Ha^{(U)} \psi = \Ha\, \varphi + i\, \GG_{+} \bm{c} - i\, \GG_{-} U \bm{c}\,.\label{eq:vNact}
		\end{gather}
	\end{theorem}

\begin{remark}[Asymptotic expansions]\label{rem:vNasy}
Consider any of the self-adjoint operators $\Ha^{(U)}$, $U \in \mathrm{U}(2)$, and let $\psi = \varphi + \GG_{+} \bm{c} + \GG_{-} U \bm{c} \in \dom\big(\Ha^{(U)}\big)$ be as in \eqref{eq:vNdom}. 
Building on the results derived in \cite{DR17} for the radial operators and recalling the asymptotic expansions of the modified Bessel functions \cite[\S 10.31]{NIST}, we infer that, as $r \to 0^+$,
\begin{equation}\label{eq:vNasy}
	\psi(r,\theta) = \tfrac{1}{\sqrt{2\pi}} \left[ \frac{v_0}{r^{\alpha}} + w_{0}\, r^{\alpha} 
	+ \left( \frac{v_{-1}}{r^{1-\alpha}} + w_{-1}\, r^{1-\alpha} \right) e^{-i \theta}
	+\, o(r) \right] ,
\end{equation}
where the coefficients $\vl,\wl \in \C$ ($\ell \in \{0,-1\}$) are given by
\begin{gather}
	\vl \doteq \frac{\Gamma(|\ell+\alpha|)}{2^{1-|\ell+\alpha|}}\, \Big( c_{\ell} + (U \bm{c})_{\ell} \Big)\,, \label{eq:defv}\\
	\wl \doteq \frac{\Gamma(-|\ell+\alpha|)}{2^{1+|\ell+\alpha|}}\, \Big( e^{- i \frac{\pi}{2}\, |\ell+\alpha|}\,c_{\ell} + e^{i \frac{\pi}{2}|\ell + \alpha|}\,(U \bm{c})_{\ell} \Big)\,.\label{eq:defw}
\end{gather}
Correspondingly, the operator domain $\dom\big(\Ha^{(U)}\big)$ can be uniquely identified as the set of functions $\psi \in \dom\big(\dHa^{*}\big)$, see \eqref{eq:domHas}, matching the asymptotics written in \eqref{eq:vNasy}-\eqref{eq:defw}.
\end{remark}

\begin{remark}[Rotation invariant and homogeneous extensions]\label{rem:vNhom}
We noticed in Remark \ref{rem:homdil} that the symmetric operator $\Ha$ is invariant under rotations and homogeneous of degree $-2$ under dilations. Making reference to the operators $T_{R}$, $R \in \mathrm{SO}(2)$, introduced in that context and considering the asymptotic expansion reported in \eqref{eq:vNasy}, it appears that the condition $T_{R}^{-1} \dom\big(\Ha^{(U)}\big) \subset \dom\big(\Ha^{(U)}\big)$ can be fulfilled only if the extension matrix $U$ is diagonal. All the other extensions actually mix the {\it s-wave} and {\it p-wave} sectors. Similarly, regarding the dilation operators $D_{\gamma}$, $\gamma \in \R$, the condition $D_{\gamma}^{-1} \dom\big(\Ha^{(U)}\big) \subset \dom\big(\Ha^{(U)}\big)$ requires either
\begin{equation}\label{eq:UFUK}
	U = \begin{pmatrix} \,-1\; & \;0\, \\ \,0\; & \;-1\, \end{pmatrix}	\qquad \mbox{or} \qquad 
	U = \begin{pmatrix} \,-e^{- i \pi \alpha}\, & \,0 \\ \,0\, & \,-e^{- i \pi (1-\alpha)}\, \end{pmatrix}.
\end{equation}
As a matter of fact, the choices of $U$ reported in \eqref{eq:UFUK} identify the only two self-adjoint extensions $\Ha^{(U)}$ which enjoy the same homogeneity of $\Ha$ under dilations. We will elaborate further on this fact in Remark \ref{rem:homFK} below. Therein, we shall show that the first and the second choice in \eqref{eq:UFUK} correspond, respectively, to the Friedrichs and Kre{\u\i}n realizations.
\end{remark}

\begin{remark}[Physical meaning of different self-adjoint realizations]\label{rem:resconv}
Some intuition on the physics described by different self-adjoint realizations of $\Ha$ can be gained making reference to appropriate regularization and limiting procedures. In fact, distinct Hamiltonians $\Ha^{(U)}$, $U \!\in\! \mathrm{U}(2)$, can be obtained as uniform resolvent limits of regular Schr\"odinger operators which only comprise smooth electric and magnetic fields, depending on the presence of zero-energy resonances \cite{Ta01}.
\end{remark}

\subsection{Boundary triplets}
Building on the results derived in the previous paragraph, we sketch hereafter the boundary triplets description of the self-adjoint realizations of $\Ha$. For the general theory of boundary triplets we refer to \cite[Ch. 14]{Scm};
see also \cite{BLLR,BM11,DHM20}. More details on the specific application to the AB model can be found in \cite{DS98} and \cite{PR11}.

Let us refer again to the symmetric operator $\dHa$ and to its adjoint $\dHa^{*}$, see \eqref{eq:domHas}. For any $\psi,\phi \in \dom\big(\dHa^{*}\big)$, a two-fold integration by parts yields
\begin{equation}\label{eq:HsHs}
	\langle \psi \,|\, \dHa^{*} \phi \rangle - \langle \dHa^{*} \psi \,|\, \phi \rangle
	= \lim_{r \to 0^{+}} r \int_{0}^{2\pi} \!\!d\theta \left[ \overline{\psi}\, (\partial_r \phi) - \overline{(\partial_r \psi)}\,\phi \right] .
\end{equation}
On the other hand, keeping in mind that $\dom\big(\dHa^{*}\big) = \dom\big(\dHa\big) \oplus \KK_{+} \oplus \KK_{-}$ and exploiting explicit asymptotic expansions of the form \eqref{eq:vNasy} close to the origin, we infer
\begin{equation*}
	\lim_{r \to 0^{+}} r \!\int_{0}^{2\pi} \!\!\!\!d\theta \left[ \overline{\psi} \,(\partial_r \phi) - \overline{(\partial_r \psi)}\, \phi \right]
	= 2 \!\sum_{\ell \in \{0,-1\}}\!\! |\ell + \alpha| \left[\overline{v_{\ell}(\psi)}\, w_{\ell}(\phi) - \overline{w_{\ell}(\psi)}\, v_{\ell}(\phi) \right],
\end{equation*}
where, in agreement with \eqref{eq:defv} and \eqref{eq:defw}, we have set
\begin{gather*}
	v_{\ell}(\psi) \doteq \lim_{r \to 0^{+}} r^{|\ell + \alpha|} \int_{0}^{2\pi} \!\!\!d\theta\, \psi(r,\theta)\,\tfrac{e^{-i \ell \theta}}{\sqrt{2\pi}}\,, \\
	w_{\ell}(\psi) \doteq \lim_{r \to 0^{+}} \tfrac{1}{r^{|\ell + \alpha|}} \left( \int_{0}^{2\pi} \!\!\!d\theta\, \psi(r,\theta)\,\tfrac{e^{-i \ell \theta}}{\sqrt{2\pi}} - \tfrac{v_{\ell}(\psi)}{r^{|\ell + \alpha|}} \right) .
\end{gather*}
Then, introducing the linear maps
\begin{gather}
\bb_1 : \dom\big(\dHa^{*}\big) \to \C^2\,, \qquad \bb_1 \psi \doteq v_{0}(\psi) \oplus v_{-1}(\psi)\,, \label{eq:b1}\\
\bb_2 : \dom\big(\dHa^{*}\big) \to \C^2\,, \qquad \bb_2 \psi \doteq 2 \alpha\, w_{0}(\psi) \oplus 2 (1-\alpha)\, w_{-1}(\psi) \,,\label{eq:b2}
\end{gather}
we may rephrase \eqref{eq:HsHs} as an abstract Green identity:
\begin{equation*}
	\langle \psi \,|\, \dHa^{*} \phi \rangle - \langle \dHa^{*} \psi \,|\, \phi \rangle
	= \langle \bb_1 \psi \,|\, \bb_2 \phi \rangle - \langle \bb_2 \psi \,|\, \bb_1 \phi \rangle\,.
\end{equation*}
Furthermore, it is easy to check that the direct sum $\bb_1 \oplus \bb_2 : \dom\big(\dHa^{*}\big) \to \C^2 \oplus \C^2$ is surjective. The above arguments prove that $(\C^2,\bb_1,\bb_2)$ is a {\it boundary triplet} for $\dHa$. 

One of the main results on boundary triplets \cite[Thm. 14.7]{Scm} allows to parametrize all self-adjoint extensions of the symmetric operator $\dHa$ in terms of self-adjoint relations on the boundary space $\C^2$.
We recall that a {\it self-adjoint relation} $\Xi$ on $\C^2$ is, by definition, a maximal symmetric closed subspace $\Xi \subset \C^2 \oplus \C^2$ (here `symmetric' means that $\langle \bm{a}_1 | \bm{b}_2 \rangle_{\C^2} = \langle \bm{a}_2 | \bm{b}_1 \rangle_{\C^2}$ for all $\bm{a}_1 \oplus \bm{a}_2, \bm{b}_1 \oplus \bm{b}_2 \in \Xi$, while `maximal' refers to the fact that there is no other closed symmetric relation strictly including $\Xi$). The above arguments ultimately entail the following \cite{DS98}.

	\begin{theorem}[Self-adjoint extensions - boundary triplets]\label{thm:btrip}
		Let $\alpha \in (0,1)$. Then, the self-adjoint extensions on $L^2(\R^2)$ of the closed symmetric operator $\dHa$ are in one-to-one  correspondence with the set of self-adjoint relations $\Xi$ on $\C^2$. Moreover, for any self-adjoint relation $\Xi$ on $\C^2$, the associated self-adjoint extension $\Ha^{(\Xi)}$ is obtained by restricting the adjoint operator $\dHa^{*}$ to the domain
		\begin{gather}
		\dom\big(\Ha^{(\Xi)}\big) = \big\{ \psi \!\in\! \dom\big(\dHa^{*}\big) \;\big|\; \bb_1 \psi \oplus \bb_2 \psi \in \Xi \big\}\,. \label{eq:btdom}
		\end{gather}
	\end{theorem}

\begin{remark}[Connecting the boundary triplets and von Neumann parametrizations]\label{rem:btvN}
Any self-adjoint relation $\Xi$ on $\C^2$ can be represented via a pair of $2 \times 2$ complex matrices $X_1,X_2 \in \mathrm{M}_{2 \times 2}(\C)$ such that
\begin{equation}\label{eq:X1X2}
	\mathrm{rank}(X_1,X_2) = 2\,, \qquad X_1\,X_2^{*} = X_2\,X_1^{*}\,.
\end{equation}
To be precise, we have
\begin{equation}
	\Xi = \big\{ \bm{b}_1 \oplus \bm{b}_2 \in \C^2 \oplus \C^2\,\big|\, X_1 \bm{b}_1 = X_2 \bm{b}_2 \big\}
	= \ran\big(X_1^{*} \oplus X_{2}^{*}\big)\,,
\end{equation}
see \cite{DS98,PR11,Po08}.
Accordingly, the condition $\bb_1 \psi \oplus \bb_2 \psi \in \Xi$ in \eqref{eq:btdom} may be expressed as
\begin{equation*}
	X_1\, \bb_1 \psi = X_2\, \bb_2 \psi \,.
\end{equation*}
Notice that the above parametrization in terms of pairs $(X_1,X_2)$ fulfilling \eqref{eq:X1X2} is uniquely determined only up to multiplication on the left by an invertible matrix. In fact, $(X_1,X_2)$ and $(X'_1,X'_2)$ represent the same self-adjoint relation $\Xi$ if and only if $X'_1 = Y\,X_1$, $X'_2 = Y\,X_2$ for some $Y \in \mathrm{GL}(2,\C)$ ($\mathrm{GL}(2,\C)$ is the general linear group of degree $2$).\\
In view of the above considerations, recalling the definitions \eqref{eq:b1}\eqref{eq:b2} of $\bb_1,\bb_2$ and the asymptotic expansion \eqref{eq:vNasy}-\eqref{eq:defw}, it can be checked by direct inspection that the von Neumann extension $\Ha^{(U)}$ and the boundary triplet analogue $\Ha^{(\Xi)}$ are related by
\begin{equation*}
	X_1 \,D_1\, ( U + \bm{1}) = X_2 \, D_2\, (U - U_{\natural})\,,
\end{equation*}
where we have set $D_1 = \big(\tfrac{\Gamma(\,|\ell+\alpha|\,)}{2^{1-|\ell+\alpha|}}\, \delta_{\ell \ell'}\big)$, $D_2 = \big(\tfrac{\Gamma(1-|\ell+\alpha|\,)}{2^{|\ell+\alpha|}}\,e^{i \frac{\pi}{2}|\ell+\alpha|}\, \delta_{\ell \ell'}\big)$ and $U_{\natural} = \big(- e^{-i \pi |\ell +\alpha|} \delta_{\ell\,\ell'}\big)$. Let us highlight two notable cases: $U = -\bm{1}$, matching $X_2 = \bm{0}$, whence $\bb_1 \psi = \bm{0}$, and $U = U_{\natural}$, matching $X_1 = \bm{0}$, whence $\bb_2 \psi = \bm{0}$. The distinguished role of the self-adjoint extensions associated to these two cases was already singled out in Remark \ref{rem:vNhom}, where we argued that they are the only realizations which preserve the same homogeneity of the basic differential operator under dilations.
\end{remark}

\begin{remark}[Parametrization by projectors and Hermitian matrices]
Another way to parametrize any self-adjoint relation $\Xi$ on $\C^2$ is to use an orthogonal projector $P$ and an Hermitian matrix $T = T^*$ on $\ran P$. In accordance, using \cite[Thm. 4.3]{Po08}, we may characterize the domain of the Hamiltonian $\Ha^{(\Xi)}$ as
\begin{equation*}
	\dom\big(\Ha^{(\Xi)}\big) = \big\{ \psi \!\in\! \dom\big(\dHa^{*}\big) \;\big|\; \bb_1 \psi \in \dom(T)\,, \;T \bb_1 \psi = P \bb_2 \psi \big\}\,.
\end{equation*}
The explicit relation between the present parametrization and the one in terms of a pair $(X_1,X_2)$ described in the preceding Remark \ref{rem:btvN} can be found in \cite[Thm. 4.5]{Po08}.
\end{remark}

\subsection{Quadratic forms}
In this paragraph we provide an alternative characterization of the self-adjoint extensions of $\Ha$, based on the construction of the associated quadratic forms. To this avail, we expand on the concepts first described in \cite{CO18}; see also \cite{Te90}.
It is worth noting that, contrary to the von Neumann construction, this approach can be easily adapted to more general contexts, where no direct access to the deficiency subspaces is available \cite{CF21,CF24a,CF24b,F24}. On top of that, it lays the ground for the Kre{\u\i}n formalism, to be presented in the next paragraph.

We begin by introducing the form associated to the symmetric operator $\Ha$, \emph{i.e.},
\begin{equation}\label{eq:Qadef}
	\Qa[\psi] \doteq \langle \psi\,|\,\Ha \psi \rangle = \|(-i\nabla + \Aa)\psi\|_{2}^2\,, \qquad
	\mbox{for\; $\psi \in C^{\infty}_c(\R^2 \setminus \{\mathbf{0}\})$}\,.
\end{equation}
It is noteworthy that, working in polar coordinates and considering the natural decomposition
\begin{equation}\label{eq:psiharm}
	\psi(r,\theta) = \sum_{\ell \,\in\, \Z} \psi_{\ell}(r)\,\tfrac{e^{i \ell \theta}}{\sqrt{2\pi}} \,,
\end{equation}
the quadratic form \eqref{eq:Qadef} can be equivalently expressed as
\begin{equation}\label{eq:QaPol}
	\Qa[\psi] = \sum_{\ell \,\in\, \Z} \, \int_{0}^{\infty}\! dr\;r \left[\, |\psi'_\ell(r)|^2 + \tfrac{(\ell + \alpha)^2}{r^2}\,|\psi_{\ell}(r)|^2 \,\right] .
\end{equation}

Now, realizing that $\|\cdot\|_{2} + \Qa[\cdot]$ does indeed constitute a norm, we proceed to define, as usual, the {\it Friedrichs realization} of $\Qa$ by
\begin{equation*}
	\dom\big[\QaF \big] \doteq\, \overline{C^{\infty}_c(\R^2 \setminus \{\mathbf{0}\})}^{\;\|\,\cdot\,\|_2 \,+\, \Qa[\,\cdot\,]}, \qquad \QaF[\psi] \doteq \Qa[\psi]\,.
\end{equation*}
It appears that $\QaF$ is closed and positive semi-definite. So, by a standard argument of functional analysis (see, \emph{e.g.}, \cite[Thm. VIII.15]{RS}), it uniquely identifies a self-adjoint operator $\HaF$, with domain
\begin{align*}
	& \dom\big(\HaF\big) = \Big\{ \psi_{2} \!\in\! \dom\big[ \QaF \big] \,\;\Big|\;\, \exists\; \varphi \equiv \HaF \psi_{2} \in L^2(\R^2)\;\;\mbox{s.t.}\; \\
	& \hspace{4.5cm} \QaF[\psi_{1},\psi_{2}] = \langle \psi_{1}\,|\, \varphi\rangle \quad \forall\,\psi_{1} \!\in\! \dom\big[\QaF\big] \Big\}\,,
\end{align*}
where $\QaF[\psi_{1},\psi_{2}] = \langle (-i \nabla +\Aa) \psi_1\,|\,(-i \nabla +\Aa) \psi_2 \rangle$ is the sesquilinear form defined by polarization starting from the Friedrichs quadratic form $\QaF$. 

Using the representation \eqref{eq:QaPol} for $\Qa$, by few additional computations we deduce the following result \cite{CF21}.

\begin{proposition}[Friedrichs realization]\label{prop:QHF}
Let $\alpha \in (0,1)$. Then, the following statements hold true.
\begin{enumerate}
\item The quadratic form $\QaF$ is closed and non-negative on its domain
	\begin{equation}\label{eq:DQaSF}
		\dom\big[\QaF\big] = \big\{\psi \!\in\! H^1(\R^2)\;\big|\; \Aa \psi \!\in\! L^2(\R^2) \big\}\,.
	\end{equation}

\item The unique self-adjoint operator $\HaF$ associated to $\QaF$ is
	\begin{equation}\label{eq:DHaSF}
		\dom\big(\HaF\big) = \big\{\psi \!\in\! D\big[\QaF\big] \;\big|\; \Ha\,\psi \!\in\! L^2(\R^2) \big\}\,, \qquad
		\HaF \psi = \Ha\, \psi .\vspace{0.15cm}
	\end{equation}
	
\end{enumerate}
\end{proposition}

\begin{remark}[Boundary conditions for the Friedrichs realization]\label{rem:asyF}
We already mentioned that the asymptotic behavior close to the flux singularity is crucial for identifying different self-adjoint extensions of $\Ha$. In this connection, we introduce the {\it angular average} for any $f \in L^1_{\mathrm{loc}}(\R^2)$:
\begin{equation*}
	\big\langle f \big\rangle(r) \doteq {1 \over 2\pi}\int_{0}^{2\pi}\!\!\! d\theta\, f(r,\theta)\,.
\end{equation*}
Exploiting again the decomposition \eqref{eq:psiharm} in angular harmonics, it is easy to check that the conditions $\Aa \psi \in L^2(\R^2)$ and $\nabla \psi \in L^2(\R^2)$ in \eqref{eq:DQaSF} imply, respectively,
	\begin{equation*}
		\lim_{r \to 0^{+}} \left\langle |\psi|^2 \right\rangle(r) = 0\,, \qquad
		\lim_{r \to 0^{+}} r^2 \left\langle |\partial_r\psi|^2 \right\rangle(r) = 0\,.
	\end{equation*}
These boundary conditions must be fulfilled, in particular, by any $\psi \in \dom\big[\QaF\big]$. Considering this and recalling what was said in Remark \ref{rem:vNasy}, we can deduce the asymptotic behavior of the functions in the Friedrichs operator domain: for any $\psi \in \dom\big(\HaF\big)$, in the limit $r \to 0^+$ there holds
\begin{equation}\label{eq:vNasyFr}
	\psi(r,\theta) = \tfrac{1}{\sqrt{2\pi}} \left[ w_{0}\, r^{\alpha} + w_{-1}\, r^{1-\alpha}\, e^{-i \theta} +\, o(r) \right] ,
\end{equation}
for some $w_{0},w_{-1} \in \C$. This shows that the Friedrichs Hamiltonian coincides with the von Neumann extension corresponding to $U = - \bm{1} \in \mathrm{U}(2)$, that is
\begin{equation}\label{eq:FrU1}
	\HaF = \Ha^{(U)} \Big|_{U \,=\, - \bm{1}}\,.
\end{equation}
\end{remark}

It is well-known that the Friedrichs extension of any given symmetric operator is the one with the smallest form domain. To obtain self-adjoint extensions of $\Ha$ different from $\HaF$, we now describe how to properly enlarge the form domain, modifying accordingly the expression of the quadratic form.

As a preliminary step, we examine the defect equation, for $\mu > 0$,
\begin{equation*}
	(\Ha + \mu^2)\, \Gl_{\mu} = 0\,, \qquad \mbox{in\; $\R^2 \setminus \{0\}$}\,.
\end{equation*}
By calculations similar to those described in the previous paragraph, working in polar coordinates it can be shown that the only solutions in $L^2(\R^2)$ of this equation are (\emph{cf.} Eqs. \eqref{eq:chiell} and \eqref{eq:defGpm})
\begin{equation}\label{eq:G01exp}
	\Gl_{\mu}^{(\ell)}(r,\theta) \doteq \mu^{|\ell+\alpha|}\,K_{|\ell+\alpha|}(\mu r)\,\tfrac{e^{i \ell \theta}}{\sqrt{2\pi}}
	\qquad\;		\big(\ell \!\in\!\{0,-1\}\big)\,. 
\end{equation}
Incidentally, we mention that
\begin{equation}
	\big\| \Gl_{\mu}^{(\ell)} \big\|_{2}^2 = \tfrac{\pi |\ell+\alpha|}{2 \sin(\pi \alpha)}\, \mu^{2|\ell + \alpha|-2}\,, \label{eq:GL2norm}\\
\end{equation}
and, as $r \to 0^{+}$,
\begin{equation}
	\Gl_{\mu}^{(\ell)}(r,\theta) = 
	\left[\tfrac{\Gamma(|\ell+\alpha|)}{2^{1-|\ell+\alpha|}}\,\tfrac{1}{r^{|\ell+\alpha|}} + \tfrac{\Gamma(-|\ell+\alpha|)}{2^{1+|\ell+\alpha|}}\,\mu^{2|\ell+\alpha|} r^{|\ell+\alpha|} + o(r)\right]\!\tfrac{e^{i \ell \theta}}{\sqrt{2\pi}}\,, \label{eq:Gasy}
\end{equation}
see \cite[Eq. 6.521.3]{GR} and \cite[\S 10.31]{NIST}. Moreover, in view of \eqref{eq:vNasyFr} and \eqref{eq:Gasy}, it follows that $\Gl_{\mu_1}^{(\ell)} - \Gl_{\mu_2}^{(\ell)} \in \dom\big(\HaF\big)$ for all $\mu_1,\mu_2 > 0$.

On account of the above considerations, we introduce trial functions of the form
\begin{equation*}
	\psi = \phi_{\mu} + \mbox{$\sum_{\ell\,\in\,\{0,-1\}}$}\,q_{\ell}\,\Gl_{\mu}^{(\ell)}\,,
\end{equation*}
with $\phi_{\mu} \!\in\! \dom\big[\QaF\big]$ and $\bm{q} = (q_{0},q_{-1}) \!\in\! \C^2$.
For such functions, a heuristic evaluation of the expectation value $\langle \psi\,|\,\Ha \psi \rangle$ suggests the following educated guess:
\begin{equation}\label{eq:QaB}
	\Qa^{(B)}[\psi] \doteq \QaF[\phi_{\mu}] + \mu^2\,\|\phi_{\mu}\|_2^{2} - \mu^2\,\|\psi\|_2^{2} + \bm{q}^{T} \big[L(\mu) + B\big]\, \bm{q}\,,
\end{equation}
where we have put
\begin{equation}\label{eq:Ldef}
	L(\mu) \doteq \tfrac{\pi}{2\sin(\pi \alpha)} \begin{pmatrix} \,\mu^{2\alpha}\; & \;0\, \\ \,0\; & \;\mu^{2(1-\alpha)} \end{pmatrix}\,,
\end{equation}
and $B \in \mathrm{M}_{2 \times 2,\mathrm{Herm}}(\C)$ is any Hermitian matrix, labeling the quadratic form \eqref{eq:QaB}.
The expression $\bm{q}^{T}[L(\mu) + B] \bm{q}$ in \eqref{eq:QaB} is fixed on purpose so that the form $\Qa^{(B)}[\psi]$ is independent of the spectral parameter $\mu$, whence of the specific decomposition employed for $\psi$.

Building on the above arguments and using again \cite[Thm. VIII.15]{RS}, we derive the following result \cite{CO18,F24}.

	\begin{theorem}[Self-adjoint extensions - quadratic forms]\label{thm:Qext}
		Let $\alpha \in (0,1)$ and $ \mu > 0 $. Then, for any $B \in \mathrm{M}_{2 \times 2,\mathrm{Herm}}(\C)$, the following statements hold true.
		\begin{enumerate}
			\item The quadratic form $\Qa^{(B)}$ is well-defined on the domain
				\begin{equation}\label{eq:QBdom}
					\dom\big[\Qa^{(B)}\big] \doteq \Big\{ \psi = \phi_{\mu} + \mbox{$\sum_{\ell \in \{0,-1\}}$}\, q_{\ell}\, \Gl_{\mu}^{(\ell)} \!\in\! L^2(\R^2) \,\Big|\, \phi_{\mu} \!\in\! \dom\big[\QaF\big],\; \bm{q} \!\in\! \C^2 \Big\}\,;
				\end{equation}
				moreover, it is independent of $\mu$, closed and bounded from below.
			
			\item The unique self-adjoint operator $\Ha^{(B)}$ associated to $\Qa^{(B)}$ is given by
				\begin{align}
					& \dom\big(\Ha^{(B)}\big) \doteq \Big\{ \psi = \phi_{\mu} + \mbox{$\sum_{\ell \in \{0,-1\}}$}\, q_{\ell} \,\Gl_{\mu}^{(\ell)} \!\in\! \dom\big[\Qa^{(B)}\big] \;\Big|\; \phi_{\mu} \!\in\! \dom\big(\HaF\big)\,, \; \bm{q} \!\in\! \C^2, \nonumber \\
					& \hspace{0.9cm} \lim_{r \to 0^{+}} \tfrac{\pi\;2^{|\ell+\alpha|}\, \Gamma(|\ell+\alpha| )}{r^{|\ell+\alpha|}} \big(|\ell+\alpha| + r\, \partial_r \big) \big\langle \phi_{\mu}\,\tfrac{e^{-i \ell \theta}}{\sqrt{2\pi}} \big\rangle = \big[\big(L(\mu) + B\big) \bm{q} \big]_{\ell} \,\bigg\}\,, \label{eq:HBdom}
				\end{align}
				\begin{equation}\label{eq:HBaction}
					\big( \Ha^{(B)} + \mu^2 \big) \psi \doteq \big( \HaF + \mu^2\big) \phi_{\mu}\,.	
				\end{equation}		
		\end{enumerate}
	\end{theorem}

\begin{remark}[The Kre{\u\i}n extension]
Taking into account that the form domain $\dom\big[\Qa^{(B)}\big]$ is the same for all $B \in \mathrm{M}_{2 \times 2,\mathrm{Herm}}(\C)$, see \eqref{eq:QBdom}, it is possible to introduce a partial order relation for the associated self-adjoint operators. Namely, $\Ha^{(B_1)}$ is said to be {\it smaller} than $\Ha^{(B_2)}$ if $\Qa^{(B_1)}[\psi] \leqslant \Qa^{(B_2)}[\psi]$ for all $\psi \in \dom\big[\Qa^{(B_1)}\big] \equiv \dom\big[\Qa^{(B_2)}\big]$. With this understanding, among all the self-adjoint operators parametrized in Theorem \ref{thm:Qext}, there exists a {\it smallest positive} one \cite[Thm. 13.12]{Scm}. This is called the Kre{\u\i}n extension and we shall indicate it with $\HaK$.
Considering that $\QaF$ and $L(\mu)$ are non-negative, by inspection of \eqref{eq:QaB} it appears that the the Kre{\u\i}n Hamiltonian corresponds to the choice $B = 0$, namely,
\begin{equation}\label{eq:KreinB}
	\HaK = \Ha^{(B)}\Big|_{B\,=\,0}\,.
\end{equation}
\end{remark}

\begin{remark}[Connecting the quadratic forms and von Neumann parametrizations]\label{rem:asyQuad}
Let $B \in \mathrm{M}_{2 \times 2,\mathrm{Herm}}(\C)$ and $\psi \in \dom\big(\Ha^{(B)}\big)$. Recalling what we said in Remark \ref{rem:asyF} and using the boundary condition in Eq. \eqref{eq:HBdom}, it follows that, as $r \to 0^{+}$,
\begin{align*}
	\psi(r,\theta) & = \sum_{\ell \in \{0,-1\}} \bigg[ 
	q_{\ell}  \frac{\Gamma(|\ell + \alpha|)}{2^{1-|\ell + \alpha|}} \,\frac{1}{r^{|\ell + \alpha|}} \\
	& \hspace{2cm} + \bigg( w_{\ell} + q_{\ell} \,\frac{\Gamma(-|\ell + \alpha|)}{2^{1+|\ell + \alpha|}}\,\mu^{2|\ell + \alpha|}\bigg) \,r^{|\ell + \alpha|}
	\bigg] \tfrac{e^{i \ell \theta}}{\sqrt{2\pi}} + o(r) \\
	& = \sum_{\ell \in \{0,-1\}} \bigg[ 
	 \frac{\Gamma(|\ell + \alpha|)}{2^{1-|\ell + \alpha|}} \,\frac{q_{\ell}}{r^{|\ell + \alpha|}} 
	 + \frac{(B\,\bm{q})_{\ell}}{2^{|\ell+\alpha|} \Gamma(|\ell+\alpha|+1)} \,r^{|\ell + \alpha|}
	\bigg] \tfrac{e^{i \ell \theta}}{\sqrt{2\pi}} + o(r)\,.
\end{align*}
Comparing this expansion with the analogue presented in Remark \ref{rem:vNasy} for the von Neumann parametrization, by simple algebraic computations we obtain the relations
\begin{equation}\label{eq:UB}
	U = - \big[ B + L\big(e^{i \frac{\pi}{4}}\big) \big]^{-1}\, [B + L\big(e^{-i \frac{\pi}{4}}\big) \big]\,,
\end{equation}
and
\begin{equation}\label{eq:BU}
	B = \big[ L\big(e^{+i \frac{\pi}{4}}\big) U + L\big(e^{-i \frac{\pi}{4}}\big) \big]\, (U + \bm{1})^{-1}\,.
\end{equation}
Here, by a slight abuse of notation, we are referring to the analytic continuation to the complex half-plane $\{\mu \in \C\,|\,\Re \mu > 0\}$ of the map $\mu \in \R_{+} \to L(\mu)$ defined in Eq. \eqref{eq:Ldef}.
The above relations \eqref{eq:UB} and \eqref{eq:BU} identify a map $U \in \mathrm{U}(2) \to B \in \mathrm{M}_{2 \times 2,\mathrm{Herm}}(\C)$, translating the von Neumann extensions $\Ha^{(U)}$ to the quadratic form Hamiltonians $\Ha^{(B)}$ and \emph{vice versa}. Notably, this map becomes singular whenever $U + \bm{1}$ has a non-trivial kernel. These singular cases can be formally recovered setting some of the entries of $B$ equal to infinity. For example, the Friedrichs Hamiltonian $\HaF$, which matches $U = -\bm{1}$ according to \eqref{eq:FrU1}, corresponds to ``$B = \infty\,\cdot\, \bm{1}$''. The meaning of this position is that the expression on the right-hand side of \eqref{eq:HBdom} indicates an arbitrary complex number whenever $\bm{q} = \bm{0}$.
Let us also notice that, in view of \eqref{eq:KreinB} and \eqref{eq:UB}, the Kre{\u\i}n Hamiltonian $\HaK$ matches the von Neumann extension with $U = \big(- e^{-i \pi |\ell +\alpha|} \delta_{\ell\,\ell'}\big)$, corresponding to the second choice in \eqref{eq:UFUK}.
\\
We finally point out that, by combining the above arguments with those reported in Remark \ref{rem:btvN}, it is possible to derive an explicit relation connecting the quadratic form parametrization of the self-adjoint realizations with the boundary triplets analogue.
\end{remark}

\begin{remark}[Homogeneity of the Friedrichs and Kre{\u\i}n extensions]\label{rem:homFK}
The results outlined in Remarks \ref{rem:homdil}, \ref{rem:vNhom}, \ref{rem:asyF} and \ref{rem:asyQuad} ensue that the only two self-adjoint extensions preserving the homogeneity of the initial symmetric operator $\Ha$ under dilations are the Friedrichs Hamiltonian $\HaF$ and the Kre{\u\i}n Hamiltonian $\HaK$.
\end{remark}

\subsection{Kre{\u\i}n's theory}
Bearing in mind the results reported in the previous section, we now proceed to describe yet another way to characterize all the self-adjoint extensions of the symmetric operator $\Ha$. More precisely, we derive hereafter the associated resolvent operators, employing some general methods \emph{\`{a} la} Kre{\u\i}n \cite{Po01,Po08}. This formulation will prove particularly well-suited for describing the spectral and scattering properties of the different AB Hamiltonians.

We firstly refer to the Friedrichs Hamiltonian $\HaF$, see Proposition \ref{prop:QHF}. Considering that it is clearly non-negative, we introduce the associated resolvent operator
\begin{equation*}
	\RaF(z) \doteq \big(\HaF - z\big)^{-1} :\, L^2(\R^2) \to \dom\big(\HaF\big)\,, \qquad
	\mbox{for\; $z \in \C \setminus [0,\infty)$}\,.
\end{equation*}
Working in polar coordinates, an explicit computation shows that $\RaF(z)$ acts in $L^2(\R^2)$ by convolution with the integral kernel \cite{AT98}
	\begin{equation}	\label{eq:Fresolv}
		\RaF(z; r,\theta; r'\!,\theta') = \sum_{\ell\in\Z}\, I_{|\ell+\alpha|}\big(\!-i\sqrt{z}\,(r\land r')\big)\,K_{|\ell+\alpha|}\big(\!-i\sqrt{z}\,(r \lor r')\big)\, \tfrac{e^{i \ell (\theta-\theta')}}{2\pi} \,.
	\end{equation}
Here and below we refer to the determination of the square root with $\Im \sqrt{z} > 0$ for all $z \in \C \setminus [0,\infty)$, which ensures, in particular, that $\Re\big(\!-i \sqrt{z}\big) > 0$. This choice is coherent with the determination implicitly used before in Eq. \eqref{eq:chiell}.

Next, in view of the boundary conditions appearing in the definition \eqref{eq:HBdom} of the self-adjoint extensions $\Ha^{(B)}$ determined by quadratic form methods, we introduce the trace operator
	\begin{gather}
		\tt \doteq \oplus_{\ell \in \{0,-1\}} \tau^{(\ell)} \,:\, \dom\big(\HaF\big) \,\to\, \C^2\,, \nonumber \vspace{0.1cm}\\
		\tau^{(\ell)} \phi \doteq \lim_{r \to 0^{+}} \frac{\pi\,2^{|\ell+\alpha|}\,\Gamma(|\ell+\alpha|)}{r^{|\ell+\alpha|}}\,\big(|\ell+\alpha| + r\, \partial_r \big) \big\langle \phi\,\tfrac{e^{-i \ell \theta}}{\sqrt{2\pi}} \big\rangle \,. \label{eq:traceop}
	\end{gather}
Recalling what we said in Remark \ref{rem:asyF} (see, in particular, Eq. \eqref{eq:vNasyFr}), it can be easily seen that $\tt$ is surjective onto $\C^2$. In the sequel, we take as a reference symmetric operator the restriction
	\begin{equation}\label{eq:Hatt}
		\Hat \doteq \HaF \!\upharpoonright \,\ker(\tt)\,,
	\end{equation}
and proceed to classify all its self-adjoint extensions, providing for each one an explicit expression of the corresponding resolvent operator.

For any $z \in \C \setminus [0,\infty)$, we use $\RaF(z)$ and $\tt$ to define the bounded operators
\begin{gather*}
	\GGc(z) \doteq \tt \,\RaF \,:\, L^2(\R^2) \,\to\, \C^2\,, \\
	\GG(z) \doteq \big(\GGc(z^*)\big)^{*} \,:\, \C^2 \,\to\, L^2(\R^2)\,.
\end{gather*}
Note that the previously mentioned surjectivity of $\tt$ implies that $\GG(z)$ is injective.
Furthermore, by means of the first resolvent identity, for all $z_0,z \in \C \setminus [0,\infty)$ we get
\begin{gather}
	\GGc(z_0) - \GGc(z) = (z_0-z)\, \GGc(z_0)\, \RaF(z)\,, \\
	\GG(z_0) - \GG(z) = (z_0-z)\, \RaF(z)\, \GG(z_0)\,. \label{eq:GzGw}
\end{gather}
The last identity shows that the range of $\ran[ \GG(z_0) - \GG(z) ] \subset  \dom\big(\HaF\big)$. So, we can apply once more the trace map and define the Weyl operator, for some $z_0 \in \C \setminus [0,\infty)$ fixed arbitrarily, as
\begin{equation}\label{eq:Lambdadef}
	\Lambda(z) \doteq \tt\left( \tfrac{\GG(z_0) + \GG(z_0^*)}{2} - \GG(z) \right) \,:\, \C^2 \,\to\, \C^2 \,.
\end{equation}
A simple computation shows that, for all $z,w \in \C \setminus [0,\infty)$,
\begin{equation*}
	\Lambda(z) - \Lambda(w) = (w-z)\,\GGc(z)\, \GG(w)\,, \qquad 
	\Lambda(z^*) = \big(\Lambda(z)\big)^{*}\,.
\end{equation*}

Without loss of generality, we henceforth fix the reference spectral parameter as
\begin{equation*}
	z_0 = -1\,.
\end{equation*}
Then, starting from Eqs. \eqref{eq:Fresolv}-\eqref{eq:Lambdadef} and exploiting some basic properties of the modified Bessel functions, a direct calculation yields the following explicit expressions (see \cite[Lemma 3.1]{BCF} for an analogous computation):
\begin{gather}
	\GG(z) \bm{p} = \sum_{\ell \in \{0,-1\}} p_{\ell}\, \Gl_{-i\sqrt{z}}^{(\ell)}\,, \qquad \mbox{for all\, $\bm{p} = (p_0,p_{-1}) \in \C^2$}\;; \label{eq:Gzexp}\\
	\Lambda(z) = \tfrac{\pi}{2\sin(\pi \alpha)} \begin{pmatrix} \, (-i\sqrt{z})^{2\alpha} - 1 \; & \;0\, \\ \,0\; & \; (-i\sqrt{z})^{2(1-\alpha)} - 1 \end{pmatrix}\,. \label{eq:Lambdaexp}
\end{gather}
Here, $z \in \C \setminus [0,\infty)$, as usual, and $\Gl_{-i\sqrt{z}}^{(\ell)}$ denotes the analytic continuation to the complex half-plane $\{\mu \in \C\,|\,\Re \mu > 0\}$ of the defect function $\Gl_{\mu}^{(\ell)}$ previously defined in \eqref{eq:G01exp} for $\mu \in \R_{+}$, that is,
\begin{equation}\label{eq:GzexpK}
	\Gl_{-i\sqrt{z}}^{(\ell)}(r,\theta) = (-i\sqrt{z})^{|\ell+\alpha|}\,K_{|\ell+\alpha|}(-i\sqrt{z}\, r)\,\tfrac{e^{i \ell \theta}}{\sqrt{2\pi}}
	\qquad\;		\big(\ell \!\in\!\{0,-1\}\big)\,. 
\end{equation}
With a similar understanding, we have
\begin{equation}\label{eq:LamL}
	\Lambda(z) = L\big(-i\sqrt{z}\big) - L(1)\,.
\end{equation}

Let now $\Pi : \C^2 \to \C^2$ be any orthogonal projector and $\Theta : \ran \Pi \to \ran \Pi$ be any self-adjoint operator. In light of \eqref{eq:Lambdaexp}, it becomes evident that the closed operator $\Theta + \Pi \Lambda(z) \Pi$ is certainly invertible for any $z \in \C \setminus [0,\infty)$ far enough from the real, positive half-line. We set
\begin{equation*}
	\Zpt \doteq \big\{z \in \C \setminus [0,\infty)\,\big|\, \exists\,\big(\Theta + \Pi \Lambda(z) \Pi\big)^{-1}\big\}\,,
\end{equation*}
and consider, for any $z \in \Zpt$, the Kre{\u\i}n-type formula
\begin{equation}\label{eq:RaPT}
	R_{\alpha}^{(\Pi,\Theta)}(z) \doteq \RaF(z) + \GG(z)\,\Pi\,\big(\Theta + \Pi \Lambda(z) \Pi\big)^{-1}\Pi\,\GGc(z)\,.
\end{equation}
Building on the previous arguments, it is not difficult to prove that $R_{\alpha}^{(\Pi,\Theta)}(z)$ is a pseudo-resolvent and further fulfills the following relations, for all $z \in \Zpt$:
\begin{gather*}
	R_{\alpha}^{(\Pi,\Theta)}(z) - R_{\alpha}^{(\Pi,\Theta)}(w) = (z-w)\,R_{\alpha}^{(\Pi,\Theta)}(z)\,R_{\alpha}^{(\Pi,\Theta)}(w)\,;\\
	R_{\alpha}^{(\Pi,\Theta)}(z^*) = \big( R_{\alpha}^{(\Pi,\Theta)}(z)\big)^{*}\,.
\end{gather*}
To say more, $R_{\alpha}^{(\Pi,\Theta)}(z)$ is injective. So, upon setting
\begin{equation}
	\Ha^{(\Pi,\Theta)} \doteq z +  \big( R_{\alpha}^{(\Pi,\Theta)}(z)\big)^{-1}
\end{equation}
an application of \cite[Thm. 2.1 and Thm. 2.4]{Po08} yields the following.

\begin{theorem}[Resolvent of the self-adjoint extensions of $\Hat$]\label{thm:HTheta}
		Let $\Pi : \C^2 \to \C^2$ be any orthogonal projector and $\Theta : \ran \Pi \to \ran \Pi$ be any self-adjoint operator. Then, $\C \setminus \R \subset \Zpt$ and the bounded linear operator $R_{\alpha}^{(\Pi,\Theta)}(z)$ defined by \eqref{eq:RaPT} is the resolvent operator for the self-adjoint extension of $\Hat$, see \eqref{eq:Hatt}, defined by 	
	\begin{gather}
			\dom\big(\Ha^{(\Pi,\Theta)}\big) \doteq \Big\{ \psi = \varphi_{z} + \GG(z) \bm{p} \in L^2(\R^2)\; \Big|\; \varphi_{z} \!\in\! \dom\big(\HaF\big),\; \bm{p} \!\in\! \ran \Pi \!\subset\! \C^2, \nonumber \\
			\hspace{6.6cm} \Pi \tt \varphi_z = \big(\Theta + \Pi \Lambda(z) \Pi \big) \bm{p} \Big\}\, , 
		\label{eq:domHT}\\
				\big(\Ha^{(\Pi,\Theta)}\! - z\big) \psi = \big(\HaF - z\big) \varphi_z\,. \label{eq:HThetadef}
		\end{gather}
		This definition is independent of $z$ and the decomposition of $\psi \in \dom\big(\Ha^{(\Pi,\Theta)}\big)$ appearing in \eqref{eq:domHT} is unique.
	\end{theorem}	

\begin{remark}[Validity of the Kre{\u\i}n formula]
		The definition \eqref{eq:RaPT} of $R_{\alpha}^{(\Pi,\Theta)}(z)$ is initially understood for $z \in \Zpt$. Yet, the same expression can be extended to any $z \in \rho\big(\HaF\big) \cap \rho\big(\Ha^{(\Pi,\Theta)}\big)$ by \cite[Thm. 2.19]{CFP18}.
\end{remark}

\begin{remark}[Connecting the Kre{\u\i}n and von Neumann parametrizations]\label{rem:KtovN}
Noting that the previously specified determination of the square root yields $-i \sqrt{\pm i} = e^{\mp i \pi/4}$, by comparison of the explicit expressions \eqref{eq:chiell}\eqref{eq:defGpm} and \eqref{eq:Gzexp}\eqref{eq:GzexpK} we infer that $G^{(\ell)}_{\pm} = \Gl_{-i\sqrt{\pm i}\,}^{(\ell)}$ and, accordingly,
\begin{equation*}
	\GG_{\pm} \equiv \GG(\pm i) \,:\, \C^2 \to \KK_{\pm} \subset L^2(\R^2)\,.
\end{equation*}
We already mentioned that these maps are in fact continuous bijections, see the comments below \eqref{eq:defGpm}. Moreover, using the identity \eqref{eq:GzGw} and recalling the definition \eqref{eq:Lambdadef} of $\Lambda(z)$, a simple algebraic computation shows that $\GG_{\pm}^{*} \GG_{\pm} = \pm i \Lambda(\mp i)$. This implies, in turn, that $\GG_{\pm}$ are unitary with respect to the modified inner product $\langle \bm{p}_1, \bm{p}_2 \rangle_{\Lambda} \doteq \big(\sqrt{\pm i \Lambda(\mp i)}\, \bm{p}_1\big)^{*} \cdot \big(\sqrt{\pm i \Lambda(\mp i)}\,\bm{p}_2\big)$ on $\C^2$.
Building on this and \cite[Thm. 4.1 and Thm. 4.3]{Po03} (see also \eqref{eq:UGGU} and \cite[Thm. 3.1]{Po08}), it can be inferred that the von Neumann and the Kre{\u\i}n parametrizations of the AB Hamiltonians given in Theorems \ref{thm:vNext} and \ref{thm:HTheta} are related by the following maps:
\begin{gather}
	U = - \,\big[ \bm{1} + 2 \big(\Theta - \Pi \Lambda(i)\Pi \big)^{-1} \Pi \Lambda(i) \Pi \big]\,; \\
	\Theta = \Pi\, \Lambda(i)\, (U - \bm{1}) (U + \bm{1})^{-1} \Pi\,.\label{eq:TU}
\end{gather}
Also in this case, a singularity occurs whenever $U + \bm{1}$ is not invertible. This pathology can be here avoided altogether by choosing the projector $\Pi$ in such a way that $\ran \Pi \subset [\ker(U + \bm{1})]^{\perp}$. In particular, the Friedrichs Hamiltonian $\HaF$ corresponds to the null projector $\Pi = \bm{0}$ ($\Theta$ becomes irrelevant in this case). With this understanding, the above relations identify a bijection between the von Neumann and the Kre{\u\i}n parametrizations. As a notable consequence, the family $\Ha^{(\Pi,\Theta)}$ comprises all admissible self-adjoint extensions of the symmetric restriction $\Hat = \HaF \!\!\upharpoonright \ker\tt$.\\
We finally notice that the above arguments, together with those described in Remarks \ref{rem:btvN} and \ref{rem:asyQuad}, allow also to relate the Kre{\u\i}n parametrization of the AB Hamiltonians with the boundary triplets and the quadratic forms analogues, respectively.
\end{remark}

\begin{remark}[Rotation invariant extensions]\label{rem:rotinv}
In view of the identity \eqref{eq:TU}, recalling the arguments reported in Remark \ref{rem:vNhom} and noting that the matrix $\Lambda(i)$ is diagonal, it appears that any Hamiltonian $\Ha^{(\Pi,\Theta)}$ which preserves the rotational symmetry of the basic differential operator $\Ha$ must necessarily match one of the following alternatives: $\Pi = \0$ is the null projector; $\Pi$ is the orthogonal projector onto the {\it s-wave} or {\it p-wave} sector, and $\Theta$ is the multiplication by any real number thereon; $\Pi = \bm{1}$ and $\Theta : \C^2 \to \C^2$ is a real diagonal matrix. 
\end{remark}

\begin{remark}[More on the connection with the quadratic form parametrization]\label{rem:conKQ}
Recalling the explicit expressions \eqref{eq:G01exp} and \eqref{eq:Gzexp}\eqref{eq:GzexpK} for $\Gl_{\mu}^{(\ell)}$ and $\GG(z)$, respectively, a direct comparison shows that the representations $\psi = \phi_{\mu} + \sum_{\ell \in \{0,-1\}} q_{\ell}\, \Gl_{\mu}^{(\ell)}$ ($\mu > 0$) and $\psi = \varphi_{z} + \sum_{\ell \in \{0,-1\}} \GG(z)\, \bm{p}$ ($z \in \C \setminus [0,\infty)$) coincide if we fix $z = -\mu^2$, $\phi_{\mu} = \varphi_{-\mu^2} \in \dom\big(\HaF\big)$ and $\bm{p} = \Pi \bm{q}$. Accordingly, making reference to the boundary relations in \eqref{eq:HBdom} and \eqref{eq:domHT}, it appears that a necessary condition for having $\dom\big(\Ha^{(B)}\big) = \dom\big(\Ha^{(\Pi,\Theta)}\big)$ is
\begin{equation*}
\Pi \big[L(\mu) + B\big] \Pi = \Theta + \Pi \Lambda(-\mu^2) \Pi\,,
\end{equation*}
which, in view of \eqref{eq:LamL}, reduces to
\begin{equation}\label{eq:BtoT}
\Theta = \Pi \big[ B + L(1)\big] \Pi\,.
\end{equation}
We already mentioned in Remark \ref{rem:KtovN} that the Friedrichs realization $\HaF$ matches the projector $\Pi = \bm{0}$. On the other side, the Kre{\u\i}n Hamiltonian $\HaK$ corresponds to fixing $\Pi = \bm{1}$ and $\Theta = L(1) = \tfrac{\pi}{2\sin(\pi \alpha)}\,\bm{1}$.
\end{remark}

\begin{remark}[Representation in terms of additive perturbations]
The trace map $\tt$, initially defined on $\dom\big(\HaF\big)$ in \eqref{eq:traceop}, does actually admit a continuous extension to $\ran \GG(z)$. In fact, recalling once more the asymptotics of the Bessel functions and exploiting a related cancellation \cite[Eq. 10.29.2]{NIST}, for any $\bm{p} \in \C^2$ we infer
\begin{align*}
	\tau^{(\ell)} \GG(z) \bm{p} 
	& = \tfrac{\Gamma(|\ell+\alpha|)}{2^{1-|\ell+\alpha|}}\,(-i\sqrt{z})^{|\ell+\alpha|}\,p_{\ell} \lim_{r \to 0^{+}} \tfrac{1}{r^{|\ell+\alpha|}} \big(|\ell+\alpha| + r\, \partial_r \big)\,K_{|\ell+\alpha|}(-i\sqrt{z}\, r) \\
	& = - \,\tfrac{\pi}{2\sin(\pi \alpha)}\,\big(\!-i\sqrt{z}\,\big)^{2|\ell+\alpha|}\,p_{\ell}\,.
\end{align*}
In other words, making again reference to the analytic continuation of the matrix $L(\mu)$, we have
\begin{equation*}
	\tt\, \GG(z) = -\, L\big(\!-i\sqrt{z}\,\big)\,.
\end{equation*}
Taking this into account and using the relations \eqref{eq:LamL}\eqref{eq:BtoT}, we may rephrase the boundary condition \eqref{eq:domHT} as
\begin{align*}
	\Pi \tt \psi = \big[\Theta + \Pi \big(\Lambda(z) - L\big(\!-i\sqrt{z}\,\big) \big) \Pi \big] \bm{p} 
	= \Pi \,B\, \Pi \bm{p} \,.
\end{align*}
Then, using the basic identity \eqref{eq:HThetadef}, recalling the definition of $\GG(z)$ and assuming that $B$ is invertible, we deduce the following chain of equalities, for any $\psi \in \dom\big(\Ha^{(\Pi,\Theta)}\big)$:
\begin{align*}
	\big(\Ha^{(\Pi,\Theta)}\! - z\big) \psi 
	& = \big(\HaF - z\big) (\psi - \GG(z) \bm{p}) \\
	& = \big(\HaF - z\big) \psi - \tt^{*}\! \bm{p} 
	= \big(\HaF - z\big) \psi - \tt^{*} \Pi\,B^{-1} \Pi\,\tt \psi \,. 
\end{align*}
This shows that, on the domain of $\Ha^{(\Pi,\Theta)}$, we have the formal identity
\begin{equation}\label{eq:addpert}
	\Ha^{(\Pi,\Theta)} = \HaF - \tt^{*} \Pi\,B^{-1} \Pi\, \tt\,.
\end{equation}
It is worth noting that the two addenda on the right-hand side of the above relation are not separately well-defined on $\dom\big(\Ha^{(\Pi,\Theta)}\big)$, yet their sum can be given a meaning by exploiting a tacit cancellation.
The identity \eqref{eq:addpert} is indeed evocative since it allows to view any operator $\Ha^{(\Pi,\Theta)}$ as a singular additive perturbation of the Friedrichs Hamiltonian $\HaF$. This, in turn, is strongly reminiscent of zero-range interactions \cite{Po01,Po03}.
In this connection, recall also what was said in Remark \ref{rem:resconv}.
\end{remark}

\section{Spectral and scattering features}
We now proceed to analyze the spectral and scattering properties of the AB Hamiltonians, characterized in the preceding section as distinct self-adjoint realizations of the basic Schr\"odinger operator $\Ha$. For definiteness, we henceforth refer exclusively to the Kre{\u\i}n's parametrization, though the same results can be derived working with anyone of the other three parametrizations described before. In particular, we refer to \cite{AT98} and to \cite{DS98,PR11} for the formulations in terms of von Neumann theory and boundary triplets, respectively.

\subsection{Basic definitions and conventions}
To avoid misunderstanding, we firstly report some basic definitions and conventions, mostly borrowed from \cite[Vol. III, \S XI]{RS}.
\\
In connection with scattering theory, we always take as a reference dynamics the one generated by the free Laplacian $\H0 \doteq -\Delta : H^2(\R^2) \to L^2(\R^2)$. Considering that the latter has purely absolutely continuous spectrum $\sigma_{\mathrm{ac}}(\H0) = [0,\infty)$, for any pair $(\Pi,\Theta)$ as in Theorem \ref{thm:HTheta}, we define the {\it wave operators} to be the strong limits
	\begin{equation*}
		\Omega^{(\Pi,\Theta)}_{\pm} \equiv \Omega_{\pm}\big(\Ha^{(\Pi,\Theta)},H_0\big) 
		\doteq \slim_{t \to \pm \infty}\; e^{i t \Ha^{(\Pi,\Theta)}} e^{-i t H_0}\,.
	\end{equation*}
If they exist, such operators are partial isometries and are said to be {\it complete} if
	\begin{equation*}
		\ran\,\Omega^{(\Pi,\Theta)}_{+} 
		= \ran \,\Omega^{(\Pi,\Theta)}_{-}
		= \ran\,P_{\mathrm{ac}}\big(\Ha^{(\Pi,\Theta)}\big) \,,
	\end{equation*}
where $ P_{\mathrm{ac}}\big(\Ha^{(\Pi,\Theta)}\big)$ is the spectral projector onto the absolute continuity subspace of $L^2(\R^2)$ associated to $\Ha^{(\Pi,\Theta)}$. The notion of {\it asymptotic completeness} requires in addition that the ranges of $\Omega^{(\Pi,\Theta)}_{\pm}$ both coincide with the orthogonal complement of bound states or, equivalently, that $\Ha^{(\Pi,\Theta)}$ has no singular continuous spectrum. 

Assuming again the existence of $\Omega^{(\Pi,\Theta)}_{\pm}$, we define the {\it scattering operator} as
	\begin{equation*}
		\mathrm{S}^{(\Pi,\Theta)} \doteq \big(\Omega^{(\Pi,\Theta)}_{+} \big)^{*}\,\Omega^{(\Pi,\Theta)}_{-}.
	\end{equation*}
By general arguments \cite[Vol. III, p. 74]{RS}, it follows that $\mathrm{S}^{(\Pi,\Theta)}$ commutes with the free Laplacian $H_0$. Therefore, $H_0$ and $\mathrm{S}^{(\Pi,\Theta)}$ can be simultaneously diagonalized. To be more precise, let $\Fou  \to L^2(\R^2)$ be the unitary Fourier transform and consider the map
\begin{equation*}
	F : L^2(\R^2) \to \!\int_{\sigma(H_0)}^{\oplus}\hspace{-0.2cm} d\mu\, \big(L^2(\R^2)\big)_{\!\lambda}\,, \quad\;
	\big(F \psi \big)_{\!\lambda\!}(\omega) \doteq \big(\Fou \psi \big)\big(\sqrt{\lambda},\omega\big) \in \big( L^2(\R^2)\big)_{\!\lambda}
\end{equation*}
where $\lambda \in [0,\infty)$, $\omega \in \mathbb{S}^1$ and $\big(L^2(\R^2)\big)_{\lambda} \equiv L^2(\mathbb{S}^{1},d\omega)$. This map provides a direct integral decomposition of $L^2(\R^2)$ with respect to the spectral measure of $H_0$. The {\it scattering matrix} is the fiber-wise restriction of $\mathrm{S}^{(\Pi,\Theta)}$ to $\big(L^2(\R^2)\big)_{\lambda}$, that is
	\begin{equation*}
		\mathrm{S}^{(\Pi,\Theta)}(\lambda): \big(L^2(\R^2)\big)_{\!\lambda}\! \to \big(L^2(\R^2)\big)_{\!\lambda}\,, \quad\;
		\mathrm{S}^{(\Pi,\Theta)}(\lambda)\, u_{\lambda} \doteq F\,\mathrm{S}^{(\Pi,\Theta)} F^{*}\, u_{\lambda}\,.
	\end{equation*}
Indicating with $\mathrm{S}^{(\Pi,\Theta)}(\lambda;\omega,\omega')$ the associated integral kernel and following \cite{Ru83,IT01}, we proceed to define the {\it scattering amplitude}
	\begin{equation*}
		a^{(\Pi,\Theta)}(\lambda;\omega,\omega') \doteq \left(\tfrac{2\pi}{i \sqrt{\lambda}}\right)^{\!\!1/2}\! \left(\mathrm{S}^{(\Pi,\Theta)}(\lambda;\omega,\omega'\big) - \delta(\omega - \omega') \right),
	\end{equation*}
and the {\it differential cross section}
	\begin{equation}\label{eq:diffcross}
		\frac{d \sigma^{(\Pi,\Theta)}}{d \omega}(\lambda,\omega) \doteq \big|a^{(\Pi,\Theta)}(\lambda;\omega,0)\big|^2\,.
	\end{equation}
In the case of Hamiltonians $\Ha^{(\Pi,\Theta)}$ invariant under rotations, see Remark \ref{rem:rotinv}, we further define the {\it phase shifts} to be those complex coefficients $\delta_{\ell}^{(\Pi,\Theta)}$ ($\ell \in \Z$) which are related to the scattering amplitude by \cite[Eq. (2.20)]{Ru83}
	\begin{equation}\label{eq:phsh}
		a^{(\Pi,\Theta)}(\lambda;\omega,\omega') = \left(\tfrac{1}{2\pi i \sqrt{\lambda}}\right)^{\!\!1/2} \sum_{\ell \in \Z} \tfrac{e^{-i\ell(\omega - \omega')}}{2\pi} \Big(e^{2i\delta_{\ell}^{(\Pi,\Theta)}} - 1\Big)\,.	
	\end{equation}

Explicit expressions for the physical quantities introduced above can be derived by standard methods of stationary scattering theory. To this purpose, let us first consider the usual plane waves
\begin{equation}\label{eq:planewave}
	f_{\k}(\x) \doteq \tfrac{1}{2\pi}\, e^{i \k \cdot \x}
		= \tfrac{1}{2\pi} \sum_{\ell \in \Z} e^{i \ell (\theta-\omega) + i \frac{\pi}{2}\,|\ell|}\, J_{|\ell|}(k\,r)\, ,
\end{equation}
where $\x = (r,\theta) \in \R_{+} \!\times\! \mathbb{S}^{1}$, $\k = (k,\omega) \in \R_{+} \!\times\! \mathbb{S}^{1}$. The second equality in \eqref{eq:planewave} follows from \cite[Eq. 8.511.4]{GR}, \cite[Eq. 10.4.1]{NIST} and it is convenient for some subsequent computations. Of course, $(f_{\k})_{\k \in \R^2}$ is a system of generalized eigenfunctions for $H_0$, fulfilling $H_{0} f_{\k} = k^2 f_{\k}$, and it can be used to express the Fourier transform as
\begin{equation}\label{eq:Fpw}
	(\Fou \psi)(\k) = \lim_{R \to + \infty} \int_{\{|\x| \leqslant R\}}\!\! d\x\; \overline{f_{\k}(\x)}\,\psi(\x)\,.
\end{equation}
Now, let $(f_{\k,\pm}^{(\Pi,\Theta)})_{\k \in \R^2}$ be distributional solutions of the eigenvalue problem
	\begin{equation}\label{eq:eigeneq}
		\Ha^{(\Pi,\Theta)} f_{\k,\pm}^{(\Pi,\Theta)} = k^2\, f_{\k,\pm}^{(\Pi,\Theta)}\,,
	\end{equation}
fulfilling the boundary condition at $\x = \0$ encoded in $\dom\big(\Ha^{(\Pi,\Theta)}\big)$, see \eqref{eq:domHT}, together with the incoming ($-$) or outgoing ($+$) Sommerfeld radiation conditions
	\begin{equation}\label{eq:somm}
		\lim_{r \to +\infty} \sqrt{r}\,\big( \partial_r \mp i k\big) \left[ f_{\k,\pm}^{(\Pi,\Theta)}(r,\theta) - f_{\k}(r,\theta) \right] = 0\,.
	\end{equation}
Using these two sets of generalized eigenfunctions for the AB Hamiltonian $\Ha^{(\Pi,\Theta)}$ and mimicking \eqref{eq:Fpw}, we define the operators
	\begin{equation}\label{eq:FouPT}
		\Fou^{(\Pi,\Theta)}_{\pm}\! : L^2(\R^2) \to L^2(\R^2)\,, \qquad
		\big(\Fou^{(\Pi,\Theta)}_{\pm} \psi\big)(\k) \doteq \lim_{R \to + \infty} \int_{\{|\x| \leqslant R\}}\!\! d\x\; \overline{f_{\k,\pm}^{(\Pi,\Theta)}(\x)}\,\psi(\x)\,.
	\end{equation}
The wave operators can then be expressed as follows (see \cite[Eq. (2.8)]{Ru83}; see also \cite[Thm. 6.2]{Ag} and \cite[Thm. 5.5]{MPS18}):
	\begin{equation}\label{eq:WopF}
		\Omega^{(\Pi,\Theta)}_{\pm} = \big(\Fou^{(\Pi,\Theta)}_{\pm}\big)^{\!*}\,\Fou\,.
	\end{equation}	
In agreement with the above identity, the generalized eigenfunctions are formally related by $f^{(\Pi,\Theta)}_{\k,\pm} = \Omega^{(\Pi,\Theta)}_{\pm} f_{\k}$. Building on this, one deduces the following distributional identity for the integral kernel associated to the scattering matrix:
	\begin{equation*}
		\mathrm{S}^{(\Pi,\Theta)}(\lambda;\omega,\omega') = \int_{\R^2}\!d\x\;\, \overline{f_{(\sqrt{\lambda},\omega),+}^{(\Pi,\Theta)}(\x)}\; f_{(\sqrt{\lambda},\omega'),-}^{(\Pi,\Theta)}(\x)\,.
	\end{equation*}
It is further important to remember that the scattering amplitude can be uniquely identified as the coefficient $a^{(\Pi,\Theta)}(k^2;\omega,\theta)$ such that, as $|\x| \to +\infty$,
	\begin{equation}\label{eq:asyamp}
		f^{(\Pi,\Theta)}_{\k,+}(\x) = \tfrac{e^{i \k \cdot \x}}{2\pi} + a^{(\Pi,\Theta)}(k^2;\omega,\theta)\,\tfrac{e^{i |\k|\,|\x|}}{|\x|^{1/2}} + o\!\left(\tfrac{1}{|\x|^{1/2}}\right) ,
	\end{equation}
where $\x = (r,\theta) \in \R_{+} \!\times\! \mathbb{S}^{1}$ and $\k = (k,\omega) \in \R_{+} \!\times\! \mathbb{S}^{1}$.

Before proceeding, let us mention that some of the arguments reported below rely on the {\it Limiting Absorption Principle} (LAP) for resolvent operators. In this connection, we refer to the spaces
	\begin{equation*}
		L^2_{u}(\R^2) \doteq L^2\big(\R^2,  (1+|\x|^2)^{u/2} d\x\big) \qquad (u \in \R)\,,
	\end{equation*}
and say that a LAP holds for $H_{\alpha}^{(\Pi,\Theta)}(z)$ if the limits
	\begin{equation*}
		R_{\alpha,\pm}^{(\Pi,\Theta)}(\lambda) \doteq \lim_{\varepsilon \to 0^{+}} R^{(\Pi,\Theta)}(\lambda \pm i \varepsilon)
	\end{equation*}
exist in the Banach space of bounded operators $\mathcal{B}\big(L^2_{u}(\R^2);L^2_{-u}(\R^2)\big)$ for some $u > 0$ and for all $\lambda \in \sigma_{\mathrm{ac}}\big(\Ha^{(\Pi,\Theta)}\big) \!\setminus\! \mathrm{e}_{+}\big(\Ha^{(\Pi,\Theta)}\big)$. Here, $\mathrm{e}_{+}\big(\Ha^{(\Pi,\Theta)}\big)$ is the (possibly empty) discrete set of eigenvalues embedded in the absolutely continuous spectrum. 
We recall that establishing LAP for some $u>0$ suffices to infer that $\Ha^{(\Pi,\Theta)}$ has no singular continuous spectrum (see, \emph{e.g.}, \cite[Thm. 6.1]{Ag} and \cite[Cor. 4.7]{MPS18}).

Finally, by a {\it zero-energy resonance} we understand a distributional solution of the equation $\Ha \psi = 0$ in $L^2_{\mathrm{loc}}(\R^2) \setminus L^2(\R^2)$, which fulfills the boundary condition at $\x = \0$ encoded in $\dom\big(\Ha^{(\Pi,\Theta)}\big)$ and remains bounded at infinity.

\subsection{The Friedrichs Hamiltonian}
We firstly consider the Friedrichs Hamiltonian $\HaF$, matching the null projector $\Pi = 0$ in the Kre{\u\i}n's parametrization (we recall once more that $\Theta$ is irrelevant in this case).
Its very definition in terms of quadratic form, see Proposition \ref{prop:QHF}, makes evident that $\HaF$ is non-negative. So, we certainly have
	\begin{equation*}
		\sigma\big(\HaF\big) \subset [0,\infty)\,.
	\end{equation*}

By decomposition in angular harmonics, it can be checked that the only distributional solutions of the eigenvalue problem $\HaF \psi_{\lambda} = \lambda\, \psi_{\lambda}$ with $\lambda \geqslant 0$ are of the form
	\begin{equation}\label{eq:psimuF}
		\psi_{\lambda}(r,\theta) = \left\{\begin{array}{ll}
			\sum_{\ell \in \Z} \big[ b_{\ell}\;r^{|\ell + \alpha|} + c_{\ell}\;r^{-|\ell + \alpha|} \big]\, \tfrac{e^{i \ell \theta}}{\sqrt{2\pi}} & \;\mbox{if\, $\lambda = 0$}\,, \vspace{0.1cm}\\
			\sum_{\ell \in \Z} \big[ d_{J,\ell}\;J_{|\ell + \alpha|}(\sqrt{\lambda}\,r) + d_{Y,\ell}\;Y_{|\ell + \alpha|}(\sqrt{\lambda}\,r) \big] \tfrac{e^{i \ell \theta}}{\sqrt{2\pi}} & \;\mbox{if\, $\lambda > 0$}\,,
		\end{array}\right.
	\end{equation}
where $b_{\ell},c_{\ell},d_{J,\ell},d_{Y,\ell}$ are suitable complex coefficients, while $J_\nu,Y_\nu$ are the ordinary Bessel functions of the first and second kind \cite[\S 10]{NIST}.

On one side, considering the large argument asymptotics of the Bessel functions \cite[\S 10.17]{NIST}, it is easy to check that there is no non-trivial choice of the coefficients making $\psi_\mu$ square-integrable in the limit $r \to +\infty$. This suffices to infer that the discrete and point spectra of $\HaF$ are indeed empty.

On the other side, given the small argument expansion of the Bessel functions \cite[\S 10.7]{NIST}, it appears that the Friedrichs asymptotics \eqref{eq:vNasyFr} for $r \to 0^{+}$ requires $c_{\ell} = 0$ and $d_{Y,\ell} = 0$. Taking this into account, it is possible to fix the coefficients $d_{J,\ell}$ so as to fulfill the Sommerfeld radiation conditions \eqref{eq:somm}. We are thus led to consider the sets of (incoming and outgoing) generalized eigenfunctions for $\HaF$ given by
	\begin{equation}\label{eq:ffF}
		f^{(\mathrm{F})}_{\k,\pm}(\x) = \tfrac{1}{2\pi}  \sum_{\ell \in \Z} e^{i \ell (\theta-\omega) + i \frac{\pi}{2} |\ell| \mp i \frac{\pi}{2}\,(|\ell + \alpha| - |\ell|)}\, J_{|\ell + \alpha|}(k\,r)\,,
	\end{equation}
where $\k = (k,\omega) \in \R_{+}\!\times\!\mathbb{S}^1$. We incidentally mention that $f^{(\mathrm{F})}_{\k,\pm} \in L^2_{-u}(\R^2)$ for any\, $u> 1$. Using $f^{(\mathrm{F})}_{\k,\pm}$ to define the modified Fourier operators $\Fou^{(\mathrm{F})}_{\pm}$ according to \eqref{eq:FouPT} and exploiting the basic relation \eqref{eq:WopF}, it was originally argued in \cite{Ru83} that the wave operators 
	\begin{equation*}
		\Omega^{(\mathrm{{F}})}_{\pm} \equiv \Omega_{\pm}\big(H_0,\HaF\big) = \Omega^{(\Pi,\Theta)}_{\pm}\big|_{\Pi \,=\, 0}
	\end{equation*}
exist and are asymptotically complete. This ensures, in particular, that the absolutely continuous spectra of $H_0$ and $\HaF$ coincide, so that $\sigma_{\mathrm{ac}}\big(\HaF\big) = [0,\infty)$, and that the singular continuous spectrum of $\HaF$ is indeed empty.

To say more, making reference to the asymptotic relation \eqref{eq:asyamp} and exploiting again the large argument asymptotics of the Bessel functions $J_\nu$, it is possible to deduce an explicit expression for the scattering amplitude, namely,
	\begin{equation}\label{eq:aF}
		a^{(\mathrm{F})}(k^2;\omega,\theta) = \sum_{\ell \in \Z} \tfrac{e^{i\ell(\theta - \omega)}}{2\pi}\;\tfrac{1}{\sqrt{2\pi i k}}\big(e^{i\pi(|\ell| - |\ell+\alpha|)} - 1\big)\,.
	\end{equation}
	
We report hereafter two major results, which can be derived by summarizing and elaborating further on the above arguments.

\begin{theorem}[Spectrum of $\HaF$]\label{thm:spectrumF}
	The Friedrichs Hamiltonian $\HaF$ has purely absolutely continuous spectrum
		\begin{equation}
			\sigma\big(\HaF\big) = \sigma_{\mathrm{ac}}\big(\HaF\big) = [0,\infty)\,.
		\end{equation}
	In particular, $\sigma_{\mathrm{pp}}\big(\HaF\big) = \sigma_{\mathrm{sc}}\big(\HaF\big) = \varnothing$.
	Moreover, there is no eigenvalue embedded in the absolutely continuous spectrum and no zero-energy resonance.
\end{theorem}

\begin{theorem}[Scattering for the pair $(H_0,\HaF)$]\label{prop: waveopF}
The wave operators $\Omega_{\pm}\big(H_0,\HaF\big)$ exist and are asymptotically complete. The scattering operator $\mathrm{S}^{(\mathrm{F})} $ exists and is unitary on $L^2(\R^2)$. 
		The integral kernel associated to the scattering matrix is
			\begin{equation}\label{eq: SFexp}
				\mathrm{S}^{(\mathrm{F})}(\lambda;\omega,\omega'\big)
				= \cos(\pi \alpha)\, \delta(\omega-\omega') + \tfrac{i}{\pi}\,\sin(\pi \alpha)\;\mathrm{P.V.}\left( \tfrac{1}{e^{i (\omega-\omega')} - 1}\right)  \,,
			\end{equation}
		where $\omega,\omega' \in \mathbb{S}^1$ and $\mathrm{P.V.}$ indicates the Cauchy principal value. Furthermore, the scattering amplitude is given by
			\begin{equation*}
				a^{(\mathrm{F})}(\lambda;\omega,\omega') = \sqrt{\tfrac{2\pi}{i \sqrt{\lambda}}} \left[ \big(\cos(\pi \alpha)- 1 \big)\, \delta(\omega-\omega') + \tfrac{i}{\pi}\,\sin(\pi \alpha)\;\mathrm{p.v.}\!\left( \tfrac{1}{e^{i (\omega-\omega')} - 1}\right)  \right] ,
			\end{equation*}
			and, for $\omega \neq 0$, the differential cross section is
			\begin{equation}\label{eq:dcsF}
				\frac{d \sigma^{(\mathrm{F})}}{d \omega}(\lambda,\omega) = 		
		\frac{1}{2\pi\sqrt{\lambda}}\; \frac{\sin^2(\pi \alpha)}{\sin^2(\omega/2)} \,.
			\end{equation}
\end{theorem}

\begin{remark}[Invariance of the scattering matrix]\label{rem:Invsc}
The explicit expression \eqref{eq: SFexp} makes evident that the scattering matrix $\mathrm{S}^{(\mathrm{F})}(\lambda)$ is indeed independent of the energy parameter $\lambda \in [0,\infty)$.
\end{remark}

\begin{remark}[The total cross section and the phase shifts]\label{rem:totcsF}
The explicit expression \eqref{eq:dcsF} for the differential cross section associated to the Friedrichs Hamiltonian $\HaF$ matches the original prediction reported in \cite{AB59}. It appears that the diffusion is maximal when the flux parameter is 
	\begin{equation*}
		\alpha = 1/2\,.
	\end{equation*}
On the other hand, the occurrence of a non-integrable singularity for $\omega \to 0^+$ (equivalently, $\omega \to 2\pi^-$), concerning scattering in the forward direction, implies that the total cross section is indeed infinite in the context under analysis, meaning that
\begin{equation*}
	\sigma^{(\mathrm{F})}(\lambda) \doteq \int_{\mathbb{S}^1} d\omega\; \frac{d \sigma^{(\mathrm{F})}}{d \omega}(\lambda,\omega) = +\infty\,.
\end{equation*}
This feature is typically associated with long-range electric potentials in scattering theory. Its occurrence in the AB model is somehow unexpected at first glance, considering that the physical magnetic field has in fact zero range. Nevertheless, it may be argued that the interaction with the ideal solenoid induces a non-trivial ``topological'' phase shift in the particle's wave function, irrespective of the distance from the solenoid itself.\\
In this connection, it is worth mentioning that the Friedrichs phase shifts $\big(\delta_{\ell}^{(\mathrm{F})}\big)_{\ell \in \Z}$, fulfilling the analogue of \eqref{eq:phsh}, can be readily deduced from \eqref{eq:aF} and read
\begin{equation*}
	\delta_{\ell}^{(\mathrm{F})} = \tfrac{\pi}{2}\big(|\ell| - |\ell+\alpha|\big) \qquad (\ell \in \Z)\,.
\end{equation*}
\end{remark}

\begin{remark}[LAP for $\HaF$]\label{rem:LAPHF}
By an argument based on Mourre-type positive commutator estimates \cite[Prop. 7.3]{IT01}, it can be proved that
		\begin{equation*}
			R^{(\mathrm{F})}_{\alpha,\pm}(\lambda) \doteq \lim_{\varepsilon \to 0^{+}} \RaF(\lambda \pm i \varepsilon) \in \mathcal{B}\big(L^2_{u}(\R^2);L^2_{-u}(\R^2)\big)\,,
		\end{equation*}
for all $\lambda \in [0,\infty)$ and for any $u > 1$, with locally uniform convergence. This LAP result provides an alternative argument for the absence of singular continuous spectrum related to the Friedrichs Hamiltonian $\HaF$. Furthermore, it lies the ground for deriving the same conclusion in connection with the other Hamiltonians $\Ha^{(\Pi,\Theta)}$.
\end{remark}

\subsection{Singular perturbations}
We now proceed to discuss the properties of the generic AB Hamiltonians $\Ha^{(\Pi,\Theta)}$. While it would be possible to perform an explicit analysis through direct calculations, retracing the arguments outlined before for the Friedrichs Hamiltonian, we prefer here to regard $\Ha^{(\Pi,\Theta)}$ as a finite-rank perturbation of $\HaF$, in resolvent sense, and refer to abstract resolvent methods. By doing so, we aim at providing a clearer and comprehensive picture.

Let us firstly highlight that, in view of the Kre{\u\i}n formula \eqref{eq:RaPT}, the difference $R_{\alpha}^{(\Pi,\Theta)}(z) - \RaF(z)$ appears to be a finite rank operator, and so, in particular, of trace class. Thus, by a straightforward application of Birman-Kuroda theorem \cite[Vol. III, Thm. XI.9]{RS}, the wave operators for the pair $(\Ha^{(\Pi,\Theta)},\HaF)$ exist and are complete. This fact, the previously mentioned results for $\Omega_{\pm}(\HaF,H_0)$ and the chain rule \cite[Vol. III, p.18, Prop. 2]{RS} ensure, in turn, the existence and completeness of the wave operators $\Omega_{\pm}^{(\Pi,\Theta)}$. As a consequence, the absolutely continuous spectra of $\Ha^{(\Pi,\Theta)}$ and $H_0$ coincide.

Asymptotic completeness can be inferred via a LAP for $R_{\alpha}^{(\Pi,\Theta)}(z)$. Referring again to the Kre{\u\i}n formula \eqref{eq:RaPT} and recalling the explicit expressions \eqref{eq:Gzexp} \eqref{eq:Lambdaexp} for $\GG(z)$ and $\Lambda(z)$, respectively, an explicit computation shows that
\begin{gather}
	\GG^{\pm}(\lambda) \doteq \lim_{\varepsilon \to 0^{+}}\GG(\lambda \pm i \varepsilon) \in \mathcal{B}\big(\C^2;L^2_{-u}(\R^2)\big)\,, \label{eq:GzLAP}\\
	\Lambda^{\pm}(\lambda) \doteq \lim_{\varepsilon \to 0^{+}} \Lambda(\lambda \pm i \varepsilon) \in \mathrm{GL}(2;\C)\,.\label{eq:LambdaLAP}
\end{gather}
for all $\lambda \in [0,\infty)$ and $u > 1$ (see \cite{BCF} for a similar result). Explicitly, we have
\begin{gather}
	\GG^{\pm}(\lambda) \bm{p} = \pm\, \tfrac{i\pi}{2} \sum_{\ell \in \{0,-1\}} p_{\ell}\; \lambda^{\!\frac{|\ell+\alpha|}{2}}\,H^{(1/2)}_{|\ell+\alpha|}\big(\sqrt{\lambda}\, r\big) \,\tfrac{e^{i \ell \theta}}{\sqrt{2\pi}}\,, \label{eq:GzexpLAP}\\
	\Lambda^{\pm}(\lambda) = \tfrac{\pi}{2\sin(\pi \alpha)} \begin{pmatrix} \, e^{\mp i \pi \alpha} \lambda^{\alpha} - 1 \; & \;0\, \\ \,0\; & \; e^{\mp i \pi (1-\alpha)} \lambda^{1-\alpha} - 1 \end{pmatrix}\,, \label{eq:LambdaexpLAP}
\end{gather}
where $\bm{p} = (p_0,p_{-1}) \in \C^2$ and $H_{\nu}^{(1)},H_{\nu}^{(2)}$ are the Hankel functions, \emph{a.k.a.} Bessel functions of the third kind \cite[\S 10.2]{NIST}.
From here and from the LAP for $\HaF$ mentioned in Remark \ref{rem:LAPHF}, it follows that
	\begin{equation*}
		R_{\alpha,\pm}^{(\Pi,\Theta)}(\lambda) \doteq \lim_{\varepsilon \to 0^{+}} R^{(\Pi,\Theta)}(\lambda \pm i \varepsilon) \in \mathcal{B}\big(L^2_{u}(\R^2);L^2_{-u}(\R^2)\big)\,,
	\end{equation*}
for all $\lambda \in [0,\infty)\setminus \mathrm{e}_{+}\big(\Ha^{(\Pi,\Theta)}\big)$ and any $u > 1$, with locally uniform convergence. This implies, in turn, the absence of singular continuous spectrum for $\Ha^{(\Pi,\Theta)}$. In addition, it can be proved by direct inspection that there are no eigenvalues embedded in the continuous spectrum, so that $\mathrm{e}_{+}\big(\Ha^{(\Pi,\Theta)}\big) = \varnothing$.

The presence of bound states below the continuous threshold can be examined by abstract resolvent methods \cite[Thm. 3.4]{Po04}, recalling once more the Kre{\u\i}n formula \eqref{eq:RaPT} for $R^{(\Pi,\Theta)}(z)$. On the other hand, zero-energy resonances can be determined by explicit computations (\emph{cf.} \cite[Prop. 2.19]{BCF}).

The results outlined above are summarized in the following theorem.

	\begin{theorem}[Spectrum of $ \Ha^{(\Pi,\Theta)} $]\label{prop: spectrumTheta}
		Let $\Pi$ be any orthogonal projector on $\C^2$ and $\Theta$ any self-adjoint operator on $\ran \Pi$. Then, for any $\alpha \in (0,1)$, the spectrum of the Hamiltonian $\Ha^{(\Pi,\Theta)}$ is given by
			\begin{equation}
				\sigma\big(\Ha^{(\Pi,\Theta)}\big) = \sigma_{\mathrm{ac}}\big(\Ha^{(\Pi,\Theta)}\big) \cup \sigma_{\mathrm{pp}}\big(\Ha^{(\Pi,\Theta)}\big)\,,
			\end{equation}
		where
			\begin{gather}
				\sigma_{\mathrm{ac}}\big(\Ha^{(\Pi,\Theta)}\big) = [0,\infty) \,, \\ 
				\sigma_{\mathrm{pp}}\big(\Ha^{(\Pi,\Theta)}\big) = \big\{\! -\mu \in \R_{-}\;\big|\, \det\big[\Theta + \Pi\Lambda(-\mu)\Pi \big] = 0 \big\} \,.
			\end{gather}
		In particular, $\sigma_{\mathrm{sc}}\big(\Ha^{(\Pi,\Theta)}\big) = \varnothing$ and there are no embedded eigenvalues.
		Moreover, the eigenfunctions associated to any negative eigenvalue $- \mu \in \sigma_{\mathrm{pp}}\big(\Ha^{(\Pi,\Theta)}\big)$ are given by $\GG(-\mu) \bm{p}$ with $ \bm{p} \in \ker \big[\Theta + \Pi \Lambda(-\mu)\Pi \big]$.
		\\
		Zero-energy resonances occur if and only if $\Theta + \Pi \Lambda(0) \Pi$ is singular. More precisely, for any $ \bm{p} \in \ker \big[\Theta + \Pi \Lambda(0) \Pi\big]$, the following is a zero-energy resonance:
		\begin{equation}
			\psi(r,\theta) = \sum_{\ell \in \{0,-1\}} p_{\ell}\,\tfrac{2^{|\ell+\alpha| - 1} \Gamma(|\ell + \alpha|)}{r^{|\ell + \alpha|}}\, \tfrac{e^{i \ell \theta}}{\sqrt{2\pi}} \,.
		\end{equation}
	\end{theorem}

\begin{remark}[Bound states for $\Ha^{(\Pi,\Theta)}$]
	The number of negative eigenvalues depends on the choice of $\Pi$ and $\Theta$, but it does not depend on the specific value of the flux parameter $\alpha \in (0,1)$ \cite{PR11}. In any case, there can be at most two bound states.
\end{remark}

Let us now elaborate further on scattering theory. The incoming and outgoing generalized eigenfunctions $(f_{\k,\pm}^{(\Pi,\Theta)})_{\k \in \R^2}$ associated to the continuous spectrum of the Hamiltonian $\Ha^{(\Pi,\Theta)}$ ca be determined by means of general results borrowed from \cite{MPS18}. In this connection, let us first mention that the trace map $\tt$, originally defined in \eqref{eq:traceop}, extends naturally to all locally smooth functions. In particular, recalling \eqref{eq:ffF}, an explicit computation gives
	\begin{gather}
		\tau^{(\ell)} f^{(\mathrm{F})}_{\k,\pm} = e^{i \frac{\pi}{2} |\ell| \mp i \frac{\pi}{2}\,(|\ell + \alpha| - |\ell|)}\,k^{|\ell + \alpha|}\,\tfrac{e^{-i \ell \omega}}{\sqrt{2\pi}}\,.
	\end{gather}
To say more, considering the explicit expression \eqref{eq:GzexpLAP} and exploiting the large argument expansion of the Hankel functions, it can be checked that
	\begin{equation*}
		\lim_{r \to +\infty} \sqrt{r}\, (\partial_r \mp i k) \big(\GG^{\pm}(\lambda) \bm{p}\big)(r,\theta) = 0\,, \qquad
		\mbox{for all $\bm{p} \in \C^2$}\,.
	\end{equation*}
Taking into account the above arguments, by a simple adaptation of \cite[Thm. 5.1]{MPS18} we introduce the functions
	\begin{equation*}
		f_{\k,\pm}^{(\Pi,\Theta)} \doteq f^{(\mathrm{F})}_{\k,\pm} + \GG^{\pm}(k^2)\, \Pi\,\big(\Theta + \Pi \Lambda^{\pm}(k^2) \Pi\big)^{-1}\Pi\, \tt f^{(\mathrm{F})}_{\k,\pm} \qquad (\k \in \R^2)\,.
	\end{equation*}
These are generalized eigenfunctions for $\Ha^{(\Pi,\Theta)}$ such that $\Ha^{(\Pi,\Theta)} f_{\k,\pm}^{(\Pi,\Theta)} = k^2 f_{\k,\pm}^{(\Pi,\Theta)}$. Besides, they fulfill the local boundary condition at $\x = \0$, as well as the incoming (-) and outgoing (+) Sommerfeld radiation conditions \eqref{eq:somm}. To say more, it is easy to check that $f_{\k,\pm}^{(\Pi,\Theta)} \in L^2_{-u}(\R^2)$ for any $u > 1$.

In agreement with \eqref{eq:asyamp}, the scattering amplitude can be deduced from the asymptotic expansion of the outgoing eigenfunction $f^{(\Pi,\Theta)}_{\k,+}$ in the limit $r \to +\infty$. Using the explicit expressions \eqref{eq:ffF} and \eqref{eq:GzexpLAP}, we obtain
	\begin{align*}
		& f_{\k,+}^{(\Pi,\Theta)}(r,\theta)= 
		\tfrac{e^{i k r \cos(\theta - \omega)}}{2\pi} 
		+ \sum_{\ell \in \Z} \tfrac{e^{i\ell(\theta - \omega)}}{2\pi}\;\tfrac{1}{\sqrt{2\pi i k}}\big(e^{i\pi(|\ell| - |\ell+\alpha|)} - 1\big)\,\tfrac{e^{i k r}}{\sqrt{r}}  \\
		& + \sqrt{\tfrac{i}{8\pi k}}\!\! \sum_{\ell,\ell' \in \{0,-1\}}\!\!\!\! (-1)^{|\ell'|} \Big[\Pi \big(\Theta \!+\! \Pi \Lambda^{+}(k^2) \Pi\big)^{-1}\!\Pi \Big]_{\ell \ell'} (- i k)^{|\ell+\alpha| + |\ell' + \alpha|}\, e^{i (\ell \theta - \ell' \omega)}\, \tfrac{e^{i k r}}{\sqrt{r}} \\
		& + o\!\left(\tfrac{1}{r^{1/2}}\right).
	\end{align*}

Summing up, we have the following.

\begin{theorem}[Scattering for the pair $\big(H_0,\Ha^{(\Pi,\Theta)}\big)$]\label{prop: waveopT}
The wave operators $\Omega_{\pm}\big(H_0,\Ha^{(\Pi,\Theta)}\big)$ exist and are asymptotically complete. The scattering operator $\mathrm{S}^{(\Pi,\Theta)} $ exists and is unitary on $L^2(\R^2)$. 
		The integral kernel associated to the scattering matrix is
			\begin{align}
		& \mathrm{S}^{(\Pi,\Theta)}(\lambda;\omega,\omega'\big) = \cos(\pi \alpha)\, \delta(\omega-\omega') + \tfrac{i}{\pi}\,\sin(\pi \alpha)\;\mathrm{p.v.}\!\left( \tfrac{1}{e^{i (\omega-\omega')} - 1}\right) \nonumber \\
				& + \tfrac{i}{4\pi}\!\! \sum_{\ell,\ell' \in \{0,-1\}}\!\!\!\! (-1)^{|\ell'|} \Big[\Pi \big(\Theta \!+\! \Pi \Lambda^{+}(\lambda) \Pi\big)^{-1}\!\Pi \Big]_{\ell \ell'} \big(- i \sqrt{\lambda}\big)^{|\ell+\alpha| + |\ell' + \alpha|}\, e^{i (\ell \omega' - \ell' \omega)},\label{eq: STexp}
			\end{align}			
		where $\omega,\omega' \in \mathbb{S}^1$ and $\mathrm{P.V.}$ indicates the Cauchy principal value. Furthermore, the scattering amplitude is
			\begin{align}
				a^{(\Pi,\Theta)}(\lambda;\omega,\omega') = \sqrt{\tfrac{2\pi}{i \sqrt{\lambda}}} \left[ \big(\cos(\pi \alpha)- 1 \big)\, \delta(\omega-\omega') + \tfrac{i}{\pi}\,\sin(\pi \alpha)\;\mathrm{p.v.}\!\left( \tfrac{1}{e^{i (\omega-\omega')} - 1}\right)  \right] \nonumber \\
				+ \sqrt{\tfrac{i}{8\pi \sqrt{\lambda}}}\!\! \sum_{\ell,\ell' \in \{0,-1\}}\!\!\!\! (-1)^{|\ell'|} \Big[\Pi \big(\Theta \!+\! \Pi \Lambda^{+}(\lambda) \Pi\big)^{-1}\!\Pi \Big]_{\ell \ell'} \big(- i \sqrt{\lambda}\big)^{|\ell+\alpha| + |\ell' + \alpha|}\, e^{i (\ell \omega' - \ell' \omega)}.\label{eq:aT}
			\end{align}
\end{theorem}

\begin{remark}[The differential and total cross sections]
The differential cross section $\frac{d \sigma^{(\Pi,\Theta)}}{d \omega}$ can be explicitly determined using the defining identity \eqref{eq:diffcross}, together with the explicit expression \eqref{eq:aT} for the scattering amplitude. We do not report here the final result for a generic choice of $(\Pi,\Theta)$, for reasons of space.
\\
Notice however that, compared to the Friedrichs case, only the {\it s-wave} and {\it p-wave} contributions are modified. This is an obvious consequence of the fact that $\Ha^{(\Pi,\Theta)}$ is a finite-rank perturbation of $\HaF$ which only affects those sectors. In particular, it is easy to see that $\frac{d \sigma^{(\Pi,\Theta)}}{d \omega}(\lambda,\omega)$ presents the same singularity as $\frac{d \sigma^{(\mathrm{F})}}{d \omega}(\lambda,\omega)$ in the forward limit $\omega \to 0$ (see Remark \ref{rem:totcsF}). This implies that the total cross section is infinite also for a generic Hamiltonian $\Ha^{(\Pi,\Theta)}$, that is 
\begin{equation*}
	\sigma^{(\Pi,\Theta)}(\lambda) \doteq \int_{\mathbb{S}^1} d\omega\; \frac{d \sigma^{(\Pi,\Theta)}}{d \omega}(\lambda,\omega) = +\infty\,.
\end{equation*}
\end{remark}

\begin{remark}[On the scattering matrix]
In contrast with the Friedrichs scenario, for a generic Hamiltonian $\Ha^{(\Pi,\Theta)}$ the scattering matrix $\mathrm{S}^{(\Pi,\Theta)}(\lambda)$ depends explicitly on the energy parameter $\lambda \in [0,\infty)$. This fact is made evident by the {\it s-wave} and {\it p-wave} additional contributions in \eqref{eq: STexp}.
\end{remark}

\begin{remark}[Hamiltonians invariant under rotations]
Whenever the Hamiltonian $\Ha^{(\Pi,\Theta)}$ of interest is invariant under rotations, the analysis becomes significantly simpler and one can further proceed to determine the phase shifts.
As a matter of fact, on account of the considerations reported in Remark \ref{rem:rotinv}, the matrix appearing in the second line of \eqref{eq:aT} turns out to be diagonal.
\\
To give an example, let us consider the Kre{\u\i}n Hamiltonian $\HaK$. We mentioned in Remark \ref{rem:conKQ} that this self-adjoint realization matches the choices $\Pi = \bm{1}$ and $\Theta = \tfrac{\pi}{2\sin(\pi \alpha)}\,\bm{1}$. Accordingly, the scattering amplitude \eqref{eq:aT} reduces to
	\begin{align*}
		a^{(\mathrm{K})}(\lambda;\omega,\omega') & =	
		\sum_{\ell \in \{0,-1\}} \tfrac{e^{i\ell(\omega' - \omega)}}{2\pi}\;\tfrac{1}{\sqrt{2\pi i \sqrt{\lambda}}}\big(e^{i\pi(|\ell| + |\ell+\alpha|)} - 1\big) \\
		& \hspace{2cm} + \sum_{\ell \in \Z \setminus \{0,-1\}} \tfrac{e^{i\ell(\omega' - \omega)}}{2\pi}\;\tfrac{1}{\sqrt{2\pi i \sqrt{\lambda}}}\big(e^{i\pi(|\ell| - |\ell+\alpha|)} - 1\big)
		\\
		& = \sqrt{\tfrac{2\pi}{i \sqrt{\lambda}}} \Big[ \big(\cos(\pi \alpha)- 1 \big)\, \delta(\omega-\omega') + \tfrac{i}{\pi}\,\sin(\pi \alpha)\;\mathrm{p.v.}\!\left( \tfrac{1}{e^{i (\omega-\omega')} - 1}\right) \nonumber \\
		& \hspace{4.8cm} - \tfrac{i}{2\pi^2}\,\sin(\pi \alpha) \left(e^{i (\omega - \omega')} - 1\right) \Big].
	\end{align*}
From here and from \eqref{eq:diffcross}, it follows that the differential cross section, for $\omega \neq 0$, is
	\begin{equation*}
		\frac{d \sigma^{(\mathrm{K})}}{d \omega}(\lambda,\omega) = 
		\tfrac{1}{\sqrt{\lambda}} 
		\left[\frac{\sin^2(\pi \alpha)}{2\pi \sin^2(\omega/2)}
		+ \frac{(2 \pi-1) \,\sin^2(\pi \alpha) \cos\omega}{\pi^3} 
		+ \frac{\sin^2(\pi \alpha)}{\pi^3}\right].
	\end{equation*}
On the other hand, the phase shifts fulfilling \eqref{eq:phsh} are given by
	\begin{align*}
		\delta_{\ell}^{(\mathrm{K})} = \left\{\!\!\begin{array}{ll}
			\tfrac{\pi}{2} \big(|\ell| - |\ell+\alpha|\big)\,, &	\qquad \mbox{for $\ell \in \Z \setminus \{0,-1\}$}\,, \vspace{0.2cm}\\
			\tfrac{\pi}{2} \big(|\ell| + |\ell+\alpha|\big)\,,	&	\qquad \mbox{for $\ell \in \{0,-1\}$}\,.
		\end{array}	\right.
	\end{align*}
\end{remark}

\begin{footnotesize}
\noindent
\textbf{Acknowledgments.}
This chapter has been partly supported by MUR grant ``Dipartimento di Eccellenza'' 2023-2027 of Dipartimento di Matematica, Politecnico di Milano.
\end{footnotesize}


\begin{thebibliography}{99.}

\bibitem[AFNN17]{AFNN17} Abatangelo, L., Felli, V., Noris, B., Nys, M.: Sharp boundary behavior of eigenvalues for Aharonov–Bohm operators with varying poles. {\it J. Funct. Anal.} {\bf 273}, 2428--2487 (2017).

\bibitem[AN18]{AN18} Abatangelo, L., Nys, M.: On multiple eigenvalues for Aharonov–Bohm operators in planar domains. {\it Nonlinear Analysis} {\bf 169}, 1--37 (2018).

\bibitem[Ab57]{Ab57} Abrikosov, A.A.: On the magnetic properties of superconductors of the second group. {\it Sov. Phys. JETP} {\bf 5}, 1174--1182 (1957).

\bibitem[AT98]{AT98} Adami, R., Teta, A.: On the Aharonov-Bohm Hamiltonian. {\it Lett. Math. Phys.} {\bf 43}, 43--54 (1998).

\bibitem[Ag75]{Ag} Agmon, S.: Spectral properties of Schr\"odinger operators and Scattering Theory. {\it Ann. Scuola Sup. Pisa} (IV) {\bf 11}, 151--218 (1975).

\bibitem[AB59]{AB59} Aharonov, Y., Bohm, D.: Significance of electromagnetic potentials in the quantum theory. {\it Phys. Rev.} {\bf 115}(3), 485--491 (1959).

\bibitem[AB61]{AB61} Aharonov, Y., Bohm, D.: Further Considerations on Electromagnetic Potentials in the Quantum Theory. {\it Phys. Rev.} {\bf 123}, 1511–1524 (1961).

\bibitem[ACR16]{ACR16} Aharonov, Y., Cohen, E., Rohrlich, D.: Nonlocality of the Aharonov-Bohm effect. {\it Phys.
Rev. A} {\bf 93}, 042110 (2016).



\bibitem[AT11]{AT11} Alexandrova, I., Tamura, H.: Resonance free regions in magnetic scattering by two solenoidal fields at large separation. {\it J. Funct. Anal.} {\bf 260}, 1836--1885 (2011).

\bibitem[AT14]{AT14} Alexandrova, I., Tamura, H.: Resonances in scattering by two magnetic fields at large separation and a complex scaling method. {\it Adv. Math.} {\bf 256}, 398--448 (2014).

\bibitem[AWH13]{AWH13} Arfken, G.B., Weber, H.J., Harris, F.E.: Mathematical Methods for Physicists: A Comprehensive Guide (Seventh Ed.). Elsevier Inc. (2013).

\bibitem[BK20]{BK20} Bartolomei, H., Kumar, M., Bisognin, R., Marguerite, A., Berroir, J.-M., Bocquillon, E., Plaçais, B., Cavanna, A., Dong, Q., Gennser, U., Jin, Y., F\`{e}ve, G.: Fractional statistics in anyon collisions. {\it Science} {\bf 368}, 173--177 (2020).

\bibitem[BT09]{BT09} Batelaan, H., Tonomura, A.: The Aharonov–Bohm effects: Variations on a subtle theme. {\it Phys. Today} {\bf 62}(9), 38--43 (2009).

\bibitem[BLLR17]{BLLR} Behrndt, J., Langer, M., Lotoreichik, V., Rohleder, J.: Quasi boundary triples and semi-bounded self-adjoint extensions. {\it Proc. Royal Soc. Edinburgh: Sec. A Mathematics} {\bf 147}(5), 895--916 (2017).

\bibitem[BM11]{BM11} Behrndt, J., Micheler, T.: Boundary triples and quasi boundary triples for elliptic operators. {\it Proc. Appl. Math. Mech.} {\bf 11}: 883--884 (2011).

\bibitem[BMO10]{BMO10} Bogomolny, E., Mashkevich, S., Ouvry, S.: Scattering on two Aharonov–Bohm vortices with opposite fluxes. {\it J. Phys. A: Math. Theor.} {\bf 43}, 354029 (2010).

\bibitem[BDELL20]{BDELL20} Bonheure, D., Dolbeault, J., Esteban, M.J., Laptev, A., Loss, M.: Symmetry results in two-dimensional inequalities for Aharonov–Bohm magnetic fields. {\it Commun. Math. Phys.} {\bf 375}, 2071--2087 (2020).

\bibitem[BCF24]{BCF} Borrelli, W., Correggi, M., Fermi, D.: Pauli Hamiltonians with an Aharonov-Bohm Flux. J. Spectr. Theory (2024). published online first (DOI: 10.4171/JST/496).

\bibitem[BPC21]{BPC21} Brosco, V., Pilozzi, L., Conti, C.: Two-flux tunable Aharonov-Bohm effect in a photonic lattice. {\it Phys. Rev. B} {\bf 104}, 024306 (2021).

\bibitem[BDG11]{BDG11} L. Bruneau, J. Derezi\'{n}ski, V. Georgescu, Homogeneous Schr\"{o}dinger Operators on Half-Line, {\it Ann. Henri Poincar\'{e}} {\bf 12}, 547--590 (2011).

\bibitem[CFP18]{CFP18} Cacciapuoti, C., Fermi, D., Posilicano, A.: On inverses of Krein's Q-functions. {\it Rend. Mat. Appl.} (7) {\bf 39}(2), 229--240 (2018).

\bibitem[Ch60]{Ch60} Chambers, R.G.: Shift of an Electron Interference Pattern by Enclosed Magnetic Flux. {\it Phys. Rev. Lett.} {\bf 5}(1), 3--5 (1960)

\bibitem[CF21]{CF21} Correggi, M., Fermi, D.: Magnetic perturbations of anyonic and Aharonov–Bohm Schr\"odinger operators. \textsl{J. Math. Phys.} \textbf{62}, 032101 (2021).

\bibitem[CF24a]{CF24a} Correggi, M., Fermi, D.: Deficiency indices for singular magnetic Schr\"odinger operators. {\it Milan J. Math.} {\bf 92}, 25--39 (2024).

\bibitem[CF24b]{CF24b} Correggi, M., Fermi, D.: Schr\"odinger operators with multiple Aharonov-Bohm fluxes. {\it Ann. Henri Poincaré} (2024).

\bibitem[CO18]{CO18} Correggi, M., Oddis, L.: Hamiltonians for Two-Anyons Systems. {\it Rend. Math. Appl.} \textbf{39}, 277--292 (2018).

\bibitem[DS98]{DS98} Dabrowski, L., \u St'\!ov\'i\v cek, P.: Aharonov-Bohm effect with $\delta$-type interaction. {\it J. Math. Phys.} {\bf 39}, 47--62 (1998).

\bibitem[DOP08]{OP08} De Oliveira, C.R., Pereira, M.: Mathematical justification of the Aharonov-Bohm hamiltonian. {\it J. Stat. Phys.} {\bf 133}, 1175--1184 (2008).

\bibitem[DF23]{DF23} J. Derezi\'nski, J. Faupin, Perturbed Bessel operators. Boundary conditions and closed realizations, {\it J. Funct. Anal.} {\bf 284}(1), 109728 (2023).

\bibitem[DFNR20]{DFNR20} J. Derezi\'{n}ski, J. Faupin, Q.N. Nguyen, S. Richard, On radial Schr\"{o}dinger operators with a Coulomb potential: General boundary conditions, {\it Adv. Oper. Theory} {\bf 5}, 1132--1192 (2020).

\bibitem[DG21]{DG21} J. Derezi\'nski, V. Georgescu, On the Domains of Bessel Operators, {\it Ann. Henri Poincar\'e} {\bf 22}, 3291--3309 (2021).

\bibitem[DR17]{DR17} Derezi\'{n}ski, J., Richard, S.: On Schr\"odinger operators with inverse square potentials on the half-line. {\it Ann. Henri Poincar\'{e}} {\bf 18}, 869--928 (2017).

\bibitem[DHM20]{DHM20} Derkach, V., Hassi, S., Malamud, M.: Generalized boundary triples, I. Some classes of isometric and unitary boundary pairs and realization problems for subclasses of Nevanlinna functions. {\it Mathematische Nachrichten} {\bf 293}, 1278--1327 (2020).

\bibitem[DFO95]{DFO95} Desbois, J., Furtlehner, C., Ouvry, S.: Random magnetic impurities and the Landau problem. {\it Nucl. Phys. B} {\bf 453}, 759--776 (1995).

\bibitem[DFO97]{DFO97} Desbois, J., Furtlehner, C., Ouvry, S.: Density correlations of magnetic impurities and disorder. {\it J. Phys. A: Math. Gen.} {\bf 30}, 7291--7300 (1997).

\bibitem[ES49]{ES49} Ehrenberg, W., Siday, R.E.: The refractive index in electron optics and the principles of dynamics. {\it Proc. Phys. Soc. Series B} \textbf{62}, 821 (1949).

\bibitem[EV02]{EV02} Erd\"os, L., Vougalter, V.: Pauli operator and Aharonov–Casher theorem for measure valued magnetic fields. {\it Commun. Math. Phys.} {\bf 225}, 399--421 (2002).

\bibitem[ESV02]{ESV02} Exner, P., \u St'\!ov\'i\v cek, P., Vyt\v ras, P.: Generalized boundary conditions for the Aharonov-Bohm effect combined with a homogeneous magnetic field. {\it J. Math. Phys.} \textbf{43}, 2151--2168 (2002).

\bibitem[FNOS23]{FNOS23} Felli, V., Noris, B., Ognibene, R., Siclari, G.: Quantitative spectral stability for Aharonov-Bohm operators with many coalescing poles. {\it arXiv:2306.05008 [math.AP]} (2023).

\bibitem[Fe24]{F24} Fermi, D.: Quadratic forms for Aharonov-Bohm Hamiltonians, pp. 205--228 in Correggi, M., Falconi M. (Eds.),
''Quantum Mathematics I'', Springer INdAM Series (SINDAMS, vol. 57), Springer Singapore (2024).

\bibitem[FG08]{FG08} Franchini, F., Goldhaber, A.S.: Aharonov–Bohm effect with many vortices. {\it Phys. Scr.} {\bf 78}, 065002 (2008).

\bibitem[GS04]{GS04} Geyler, V.A., \u St'\!ov\'i\v cek, P.: Zero modes in a system of Aharonov-Bohm fluxes. {\it Rev. Math. Phys.} {\bf 16}, 851--907 (2004).

\bibitem[GR07]{GR} Gradshteyn, I.S., Ryzhik, I.M.: Table of Integrals, Series, and Products, 7th Ed.. Academic Press, Elsevier Inc. (2007).

\bibitem[GK14]{GK14} Grillo, G., Kov\v ar\'ik, H.: Weighted dispersive estimates for two-dimensional Schr\"odinger operators with Aharonov–Bohm magnetic field. {\it J. Differ. Equ.} {\bf 256}, 3889--3911 (2014).

\bibitem[IT01]{IT01} Ito, H.T., Tamura, H.: Aharonov–Bohm effect in scattering by point-like magnetic fields at large separation. {\it Ann. Henri Poincar\'e} {\bf 2}, 309--359 (2001).

\bibitem[Ka15]{K15} Kang, K.: Locality of the Aharonov-Bohm-Casher effect. {\it Phys. Rev. A} {\bf 91}, 052116 (2015).

\bibitem[Ka22]{K22} Kang, K.: Gauge invariance of the local phase in the Aharonov-Bohm interference: quantum electrodynamic approach. {\it Europhysics Letters} {\bf 140}(4), 46001 (2022).

\bibitem[KA14]{KA14} Kenneth, O., Avron, J.E.: Braiding fluxes in Pauli Hamiltonian. {\it Ann. Phys.} {\bf 349}, 325--349 (2014).

\bibitem[KW94]{KW94} Kiers, K., Weiss, N.: Scattering from a two-dimensional array of flux tubes: A study of the validity of mean field theory. {\it Phys. Rev. D} {\bf 49}, 2081 (1994).

\bibitem[LW99]{LW99} Laptev, A., Weidl, T.: Hardy inequalities for magnetic Dirichlet forms. {\it Oper. Theory Adv. Appl.} {\bf 108}, 299--305 (1999).

\bibitem[LM77]{LM77} Leinaas, J.M., Myrheim, J.: On the theory of identical particles. {\it Nuovo Cimento B} {\bf 37}, 1--23 (1977).

\bibitem[MVG95]{MVG95} Magni, C., Valz‐Gris, F.: Can elementary quantum mechanics explain the Aharonov–Bohm effect?. {\it J. Math. Phys.} {\bf 36}, 177--186 (1995).

\bibitem[MPS18]{MPS18} Mantile, A., Posilicano, A., Sini, M.: Limiting absorption principle, generalized eigenfunctions and scattering matrix for Laplace operators with boundary conditions on hypersurfaces. {\it J. Spectr. Theory} {\bf 8}(4), 1443--1486 (2018).

\bibitem[MV20]{MV20} Marletto, C., Vedral, V.: Aharonov-Bohm phase is locally generated like all other quantum phases. {\it Phys. Rev. Lett.} {\bf 125}, 040401 (2020).

\bibitem[MOR04]{MOR04} Melgaard, M., Ouhabaz, E.-M., Rozenblum, G.: Negative discrete spectrum of perturbed multivortex Aharonov-Bohm Hamiltonians. {\it Ann. Henri Poincar\'e} {\bf 5}, 979--1012 (2004).

\bibitem[MVV22]{MVV22} Mosseri, R., Vogeler, R., Vidal, J.: Aharonov-Bohm cages, flat bands, and gap labeling in hyperbolic tilings. {\it Phys. Rev. B} {\bf 106}, 155120 (2022).

\bibitem[NLGM20]{NL20} Nakamura, J., Liang, S., Gardner, G.C., Manfra, M.J.: Direct observation of anyonic braiding statistics. {\it Nature Phys.} {\bf 16}, 931--936 (2020).

\bibitem[Na00]{Na00} Nambu, Y.: The Aharonov–Bohm problem revisited. {\it Nuclear Physics B} {\bf 579}, 590--616 (2000).

\bibitem[OLBC10]{NIST} Olver, F.W.J., Lozier, D.W., Boisvert, R.F., Clark, C.W.: NIST Handbook of mathematical functions. Cambridge University Press, Cambridge (2010).

\bibitem[PR11]{PR11} Pankrashkin, K., Richard, S.: Spectral and scattering theory for the Aharonov-Bohm operators. {\it Rev. Math. Phys.} {\bf 23}(01), 53--81 (2011).

\bibitem[Pe05]{Pe05} Perrson, M.: On the Aharonov-Casher formula for different self-adjoint extensions of the Pauli operator with singular magnetic field. {\it Electron. J. Differential Equations} {\bf 2005}, No. 55, 1--16 (2005).

\bibitem[Pe06]{Pe06} Perrson, M.: On the Dirac and Pauli operators with several Aharonov–Bohm solenoids. {\it Lett. Math. Phys.} {\bf 78}, 139--156 (2006).

\bibitem[PT89]{PT89} Peshkin, M., Tonomura, A.: The Aharonov-Bohm effect. Lecture notes in physics. Springer-Verlag, Berlin (1989).

\bibitem[Po01]{Po01} Posilicano, A.: A Krein-like Formula for Singular Perturbations of Self-Adjoint Operators and Applications. {\it J. Funct. Anal.} {\bf 183}, 109--147 (2001).

\bibitem[Po03]{Po03} Posilicano, A.: Self-adjoint extensions by additive perturbations. {\sl Annali della Scuola Normale Superiore di Pisa - Classe di Scienze}, Serie 5, Vol. 2, no. 1, 1--20 (2003).

\bibitem[Po04]{Po04} Posilicano, A.: Boundary triples and Weyl functions for singular perturbations of self-adjoint operators. {\it MFAT} {\bf 10}(2004), 57-63.

\bibitem[Po08]{Po08} Posilicano, A.: Self-adjoint extensions of restrictions. {\it Oper. Matrices} {\bf 2}, 483--506 (2008).

\bibitem[RS81]{RS} Reed, M., Simon, B.: Methods of modern mathematical physics. Academic Press (1981).

\bibitem[RY02]{RY02} Roux, P., Yafaev, D.: On the mathematical theory of the Aharonov–Bohm effect. {\it J. Phys. A: Math. Gen.} {\bf 35}, 7481--7492 (2002).

\bibitem[Ru83]{Ru83} Ruijsenaars, S.N.M.: The Aharonov-Bohm Effect and scattering theory. {\it Ann. Phys.} {\bf 146}, 1--34 (1983).

\bibitem[Sc12]{Scm} Schm\"udgen, K: Unbounded Self-adjoint Operators on Hilbert Space. Springer Dordrecht (2012).

\bibitem[St89]{St89} \v{S}\u{t}ov\'{i}\v{c}ek, P.: The green function for the two-solenoid Aharonov-Bohm effect. {\it Phys. Lett. A} {\bf 142}, 5-10 (1989).

\bibitem[St91]{St91} \v{S}\u{t}ov\'{i}\v{c}ek, P.: Krein's formula approach to the multisolenoid Aharonov-Bohm effect. {\it J. Math. Phys.} {\bf 32}, 2114--2122 (1991).

\bibitem[Ta01]{Ta01} Tamura, H.: Norm resolvent convergence to magnetic Schr\"odinger operators with point interactions. {\it Rev. Math. Phys.} \textbf{13}, 465--511 (2001).

\bibitem[Ta03]{Ta03} Tamura, H.: Resolvent convergence in norm for Dirac operator with Aharonov–Bohm field. {\it J. Math. Phys.} {\bf 44}, 2967--2993 (2003).

\bibitem[Ta07]{Ta07} Tamura, H.: Semiclassical analysis for magnetic scattering by two solenoidal fields: total cross sections.  {\it Ann. Henri Poincar\'e} {\bf 8}, 1071--1114 (2007).

\bibitem[Ta08]{Ta08} Tamura, H.: Time delay in scattering by potentials and by magnetic fields with two supports at large separation. {\it J. Funct. Anal.} {\bf 254}, 1735--1775 (2008).

\bibitem[Te90]{Te90} Teta, A.: Quadratic forms for singular perturbations of the Laplacian. \textsl{Publ. Res. Inst. Math. Sci.} \textbf{26}(5), 803--817 (1990).

\bibitem[TOMKEYY86]{TO86} Tonomura, A., Osakabe, N., Matsuda, T., Kawasaki, T., Endo, J., Yano, S., Yamada, H.: Evidence for Aharonov-Bohm effect with magnetic field completely shielded from electron wave. {\it Phys. Rev. Lett.} \textbf{56}, 792--795 (1986).

\bibitem[Va12]{V12} Vaidman, L.: Role of potentials in the Aharonov-Bohm effect. {\it Phys. Rev. A} {\bf 86}, 040101(R) (2012).

\bibitem[Wi82]{Wi82} Wilczek, F.: Quantum mechanics of fractional-spin particles. {\it Phys. Rev. Lett.} {\bf 49}, 957--959 (1982).

\bibitem[Ya03]{Ya03} Yafaev, D.: Scattering matrix for magnetic potentials with Coulomb decay at infinity. {\it Int. Eq. Op. Theory} {\bf 47}, 217--249 (2003).

\bibitem[Ya06]{Ya06} Yafaev, D.: Scattering by magnetic fields. {\it St. Petersburg Math. J.} {\bf 17}(5), 875--895 (2006).

\bibitem[Y21]{Y21} Yang, M.: Diffraction of the Aharonov–Bohm Hamiltonian. {\it Ann. Henri Poincar\'e} {\bf 22}, 3619--3640 (2021).

\end{thebibliography}
\end{document}